\documentclass[sn-maths,Numbered]{sn-jnl}

\usepackage{graphicx}%
\usepackage{multirow}%
\usepackage{amsmath,amssymb,amsfonts}%
\usepackage{amsthm}%
\usepackage{mathrsfs}%
\usepackage[title]{appendix}%
\usepackage{xcolor}%
\usepackage{textcomp}%
\usepackage{manyfoot}%
\usepackage{booktabs}%
\usepackage{algorithm}%
\usepackage{algorithmicx}%
\usepackage{algpseudocode}%
\usepackage{subcaption}
\usepackage{listings}%
\usepackage{cite}%

\theoremstyle{thmstyleone}%
\theoremstyle{thmstyletwo}%

\theoremstyle{thmstylethree}%

\begin{document}

\title[Article Title]{Complex network community discovery using fast local move iterated greedy algorithm}


\author[1,2]{\fnm{Salaheddine} \sur{TAIBI}}\email{salaheddine.taibi@univ-setif.dz}

\author*[1,2]{\fnm{Lyazid} \sur{TOUMI}}\email{lyazid.toumi@gmail.com, lyazid.toumi@univ-setif.dz}

\author[1,2]{\fnm{Salim} \sur{BOUAMAMA}}\email{salim.bouamama@univ-setif.dz}

\affil[1]{\orgdiv{Computer Science Department}, \orgname{Université Ferhat Abbas Sétif 1}, \orgaddress{\street{Algiers road}, \city{Sétif}, \postcode{19000}, \state{Setif}, \country{Algeria}}}

\affil[2]{\orgdiv{Mechatronics Laboratory (LMETR) - E1764200}, \orgname{Université Ferhat Abbas Sétif 1}, \orgaddress{\street{Algiers road}, \city{Sétif}, \postcode{19000}, \state{Setif}, \country{Algeria}}}

\abstract{\textcolor{blue}{Examining the community structures within intricate networks is crucial for comprehending their intrinsic dynamics and functionality. The paper presents the Fast Local Move Iterated Greedy (FLMIG) algorithm, a novel method designed to effectively identify community structures in intricate networks. The FLMIG algorithm improves the modularity optimization process by including a rapid local move heuristic and an iterated greedy mechanism that switches between destructive and constructive phases to strengthen the community partitions. The main innovation is the integration of random neighbor moves with an enhanced Prune Louvain algorithm, which guarantees fast convergence while maintaining the connection of the identified communities. The results of our comprehensive studies, conducted on both synthetic and and real-world networks, clearly show that FLMIG surpasses existing cutting-edge techniques in terms of both accuracy and computing efficiency. This algorithm not only provides a strong tool for identifying communities, but also makes a valuable contribution to the broader field of network analysis by offering a method that can effectively handle large-scale and dynamically evolving networks. }}

\keywords{Iterated greedy,  Community discovery, Modularity maximization, Fast local move }



\maketitle

\section{Introduction}\label{sec1}
Complex networks, which serve as mathematical representations of diverse biological, social, and technological systems, are commonly depicted as nodes linked by edges\cite{bib1}. These networks display some essential traits that are necessary for comprehending their actions. Their community structure, characterized by the formation of dense clusters called communities, is a crucial aspect for studying network dynamics. Consequently, there has been a surge in study interest regarding the identification of community structure inside intricate networks. The significance of this subject spans various disciplines, encompassing biology\cite{bib2}, energy\cite{bib3}, social sciences\cite{bib4}, personalized recommendation systems\cite{bib5}, security\cite{bib6}, epidemic spread modeling\cite{bib7}, and transportation systems\cite{bib8}.

The pursuit of efficient community identification in these networks poses multiple obstacles. These tasks involve creating strong and scalable approaches for networks with many dimensions, modifying detection techniques for networks that are either unchanging or changing over time, and maintaining important characteristics in simpler network models. Hence, the key objective of study is to develop a resilient and expandable algorithm for community detection.  Modularity is a commonly used measure for detecting community structures \cite{bib9}.  It is applied in various algorithms such as hierarchical agglomeration \cite{bib10}, the Louvain algorithm \cite{bib11}, and the \textcolor{blue}{Leiden algorithm \cite{bib12}}.  Additionally,  metaheuristics like genetic algorithms\cite{bib13}, discrete particle swarm optimization \cite{bib14}, and the bat algorithm \cite{bib15} are also utilized. These techniques prioritize modularity but frequently lack effectiveness in handling large-scale networks.

 \textcolor{blue}{ Various well-known algorithms attempt to solve these problems, with differing degrees of success. Communities in social networks may be effectively identified using the Clustering Coefficient-based Genetic Algorithm (CC-GA), while it has limitations with regard to scalability and computing complexity. The very effective Louvain method enhances modularity and uncovers hierarchical structures, although its resolution constraints may cause it to overlook tiny communities. It is dependent on the sequence in which nodes are processed. In comparison to conventional techniques, the Iterated Greedy (IG) algorithm increases modularity more quickly and accurately. However, it cannot guarantee optimum solutions and may have difficulties with highly big or complicated networks. Concurrently, the Leiden method and the Iterated Carousel Greedy (ICG) algorithm provide gains in robustness and connection, respectively, but need substantial processing power and meticulous parameter adjustment.}
 
 We provide improvements to the Louvain method in this study, with particular attention on the fast local move (FLM)  \cite{bib24} and random neighbors move techniques \cite{bib23}. To address disconnected communities, we suggest incorporating these components into the IG algorithm along with a unique refining approach to strike a balance between intensification and variety. Rapid convergence and high-quality output in vast networks are the goals of our technique.

The main contributions of this study are:

   \begin{itemize}
   \item  The proposal of the Fast Local Move Iterated Greedy algorithm (FLMIG) for community detection.
    \item The integration of the random neighbors move in the reconstruction process.
    \item The application of the novel prune algorithm within the IG framework.
    \item Demonstrating FLMIG's effectiveness and efficiency through experiments on real-world and synthetic networks.
   \end{itemize}

\textcolor{blue}{The rest of this paper is organized} in the following manner: Section \ref{sec2} examines the definitions and related works on community detection. Section \ref{sec3} provides a comprehensive explanation of the proposed FLMIG. Section \ref{sec4} provides empirical experimentation and comparative evaluations. Section \ref{sec5} finishes with conclusive remarks.

\section{Related work}\label{sec2}
\textcolor{blue}{Modularity has been proposed as a measure to evaluate the quality of detected community in networks}\cite{bib9}, an issue that is recognized as a computationally challenging optimization problem. This metric seeks to optimize the difference between the observed number of edges inside the community and the predicted number of edges. Modularity ratings that are significantly high indicate the presence of clearly established community structures.

\subsection{The problem definition}\label{subsec1}
\textcolor{blue}{A problem instance of the modularity optmization probelm can be modoled as undirected Graph}
${G = (V,E)}$, where $V$ is set of $n$ vertices and $E$ is set of $m$ edges. 

Community detection is to find a partition $P =\{C _{1} ,C _{2},....C _{k} \}$ where $V(G)= \{C _{1}  \cup   C_{2}  \cup   C_{3} ,.... \cup   C_{k}$\} and $C_{i} \cap C_{j} = \phi $.

The goal is to maximise the modularity function \cite{bib9}; the modularity function is formally defined as follows : \\
      	
\begin{equation}
      		\large Q=\frac{1}{2m}\Sigma_{i,j}^n[A_{ij}-\rho \frac{d_id_j}{2m}]\delta(c_i,c_j) \label{eq1}
\end{equation}
\textcolor{blue}{We denote $A$ as the adjacency matrix of $G$. $A_{i,j} $ = 1, there is an edge between vertex $i$ and vertex $j$; otherwise, it is 0. Moreover, $ \large\rho $ is a resolution parameter \cite{bib60} where $ \large\rho $ $ > $ 0; a higher resolution parameter guides to more communities, while a lower parameter guides to fewer communities. In our experiments, we use the traditional modularity by setting  $ \large\rho $  = 1.
}

Finally, we denote by $ d_{i} $ the degree of a vertex $i$, which is defined as: 
\begin{equation}
             \large d_{i} = \Sigma_{i}^n A_{ij} \label{eq2}
\end{equation}

Various methods have been \textcolor{blue}{developed} to tackle the issue of community detection, and they can be classified into five main types: node-centric, group-centric, network-centric, hierarchy-centric, and deep-learning approaches\cite{bib25}\cite{bib26}. Network-centric techniques primarily concentrate on studying the comprehensive structure of the network, aiming to optimize specific criteria for effectively identifying community structures. This category includes techniques that rely on modularity optimization, latent space models\cite{bib27}, Block model approximations\cite{bib25}, spectral clustering\cite{bib28}, and multi-objective optimization\cite{bib29}. In addition, approaches that focus on hierarchy can be classified into two subcategories: divisive hierarchical clustering \cite{bib30}, and agglomerative hierarchical clustering \cite{bib31}. Modularity maximization is a widely used strategy in this discipline to assess the quality of identified community partitions.

\subsection{ Modularity optimization-based methods }\label{subsec2}
\subsubsection{ Greedy methods }\label{subsubsec1}
In the last twenty years, several avaricious heuristic techniques have been created to maximize the modularity function for detecting communities in intricate networks. Newman's initial proposition was an agglomerative hierarchical clustering algorithm formulated to amalgamate communities in order to optimize the increase of modularity\cite{bib30}. This method outperformed the previous Givran and Newman algorithm\cite{bib31} in identifying communities, especially in bigger networks. Although it was effective, its efficiency was reduced by the sparse adjacency and $\Delta Q$ matrices, resulting in increased processing expenses. Clauset et al.\cite{bib32} improved the efficiency of the fast Newman method by employing data structures such as max-heaps to optimize the organization and storage of modularity variations. This enhancement resulted in a faster and more efficient algorithm.

The Louvain algorithm\cite{bib11} is a hierarchical method for optimizing modularity. It starts with a community consisting of a single node and then merges communities by locally relocating nodes to maximize the modularity function. This process persists, with networks being gradually diminished, until no more augmentation in modularity is detected. Although the Louvain algorithm is effective in big networks, it occasionally produces communities that are weakly connected\cite{bib12}. Waltman et al.  \cite{bib33} \textcolor{blue}{enhanced} this approach by incorporating an intelligent local move heuristic. This entails generating sub-networks from the initial community structure to facilitate subsequent community detection, resulting in increased performance for large networks.

V.A. Traag\cite{bib23} suggested improvements to the Louvain algorithm to boost its efficiency and decrease processing time. This involved the implementation of a stochastic neighbor relocation mechanism, where nodes are relocated to nearby communities selected at random. This modification enhances the algorithm's exploratory nature and reduces its inflexibility. Nevertheless, this alteration may marginally diminish the caliber of the answer.

In addition, Ozaki et al.\cite{bib24} proposed the Louvain prune method, which is an improvement that utilizes quick local moves to significantly reduce runtime by up to 90\% compared to the original algorithm. This variant computes modularity changes specifically for nodes that are prone to switching communities, hence minimizing superfluous repeats. Although it enhanced velocity, it failed to resolve the problem of inadequately linked neighborhoods.

In order to address this issue, V.A Traag et al.\cite{bib12} have recently devised the Leiden method. This approach combines the efficient local move, intelligent local move, and random neighbor move techniques from prior improvements made to the Louvain algorithm. The approach consists of three distinct phases: a local move heuristic, partition refinement, and network condensation based on the refined partition. These improvements have resulted in enhanced performance and preserved the quality of the designated community structures. 
             
\subsubsection{ Metaheuristic methods}\label{subsubsec2}

Metaheuristic methods have been extensively employed to discover community structures in networks through modularity optimization. These methods contrast with greedy heuristics since they commence with a preexisting answer and progressively improve upon it over iterations. The efficacy of metaheuristics depends on achieving a harmonious equilibrium between intensification and diversification. Adopting a well-balanced strategy enables more efficient exploration of the search space and the attainment of high-quality solutions\cite{bib16}. Several metaheuristic techniques have been suggested in this domain. Guimerà et al. employed simulated annealing, which incorporates both local and global node movements, to delineate community structures\cite{bib1,bib34}. Lü et al. devised an iterative tabu search method with a post-refinement procedure to maximize modularity\cite{bib35}, whereas C Liu et al. formulated an iterative local search algorithm, utilizing perturbations and local search based on label propagation to optimize the objective function\cite{bib36}.

Shi et al. utilized a genetic algorithm to perform community discovery. They employed locus-based adjacency for encoding chromosomes and applied crossover and mutation operators to improve modularity\cite{bib37}. S Bilal et al. devised an evolutionary algorithm that combines a genetic algorithm with a merging process in order to enhance the basic community structures\cite{bib38}. Said et al. introduced the clustering coefficient genetic algorithm (CC-GA), which use the highest clustering coefficient to select a neighboring chromosome for representation. Additionally, mutation extensions are used to enhance modularity\cite{bib39}.  Antonio G et al. \cite{bib40} proposed a new bio-inspired metaheuristic called the random search immune algorithm.

The bat algorithm was originally developed for addressing continuous optimization problems within the field of bio-inspired methods\cite{bib41}. Song et al. developed a discrete bat algorithm (DBA) for community detection, as described in their study\cite{bib15}. This algorithm was further improved by Kirti Aggarwal et al. \cite{bib43} by introducing a new formula for updating positions based on past experiences. Similarly, Cao et al. devised a sophisticated discrete particle swarm optimization algorithm that incorporates novel individual representation and update formula. This algorithm also incorporates a community correction strategy to enhance accuracy\cite{bib14}.

It is crucial to acknowledge that these metaheuristic methods have predominantly been evaluated on smaller and moderate sizes networks. Their utilization in extensive networks is sometimes restricted by the substantial computing time needed to attain optimal solution, hence limiting their scalability and efficacy.

\section{ Proposed approach}\label{sec3}
The Fast Local Move Iterated Greedy (FLMIG) algorithm improves upon the existing iterated greedy metaheuristic. The process begins with an initial solution generated using a local move heuristic. This solution is then improved through two main phases: first, the destruction phase, where specific parts of the existing solution are deliberately removed; then, the reconstruction phase uses a greedy heuristic to rebuild a complete solution from the segmented parts.

The FLMIG is a further development of the prior improvements made to the Louvain algorithm. During its reconstruction phase, it integrates the random neighbor move \cite{bib23} and the prune Louvain method \cite{bib24}, and combining them with the fast local move heuristic within the IG framework.

\begin{figure}[t]
  \begin{minipage}{\textwidth}
    \centering
    \begin{subfigure}[b]{.4\textwidth}
      \centering
    \includegraphics[scale=.32]{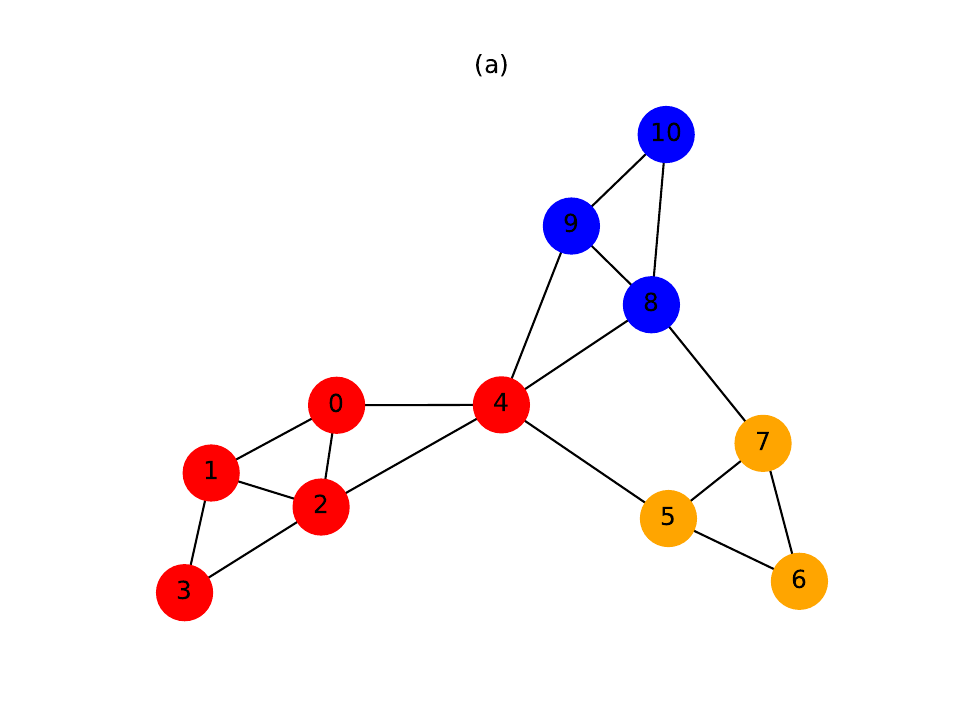} \quad
    \caption{The initialization phase}
      \label{fig:sub1a}
    \end{subfigure}\quad
    \centering
    \begin{subfigure}[b]{.4\textwidth}
      \centering
    \includegraphics[scale=.32]{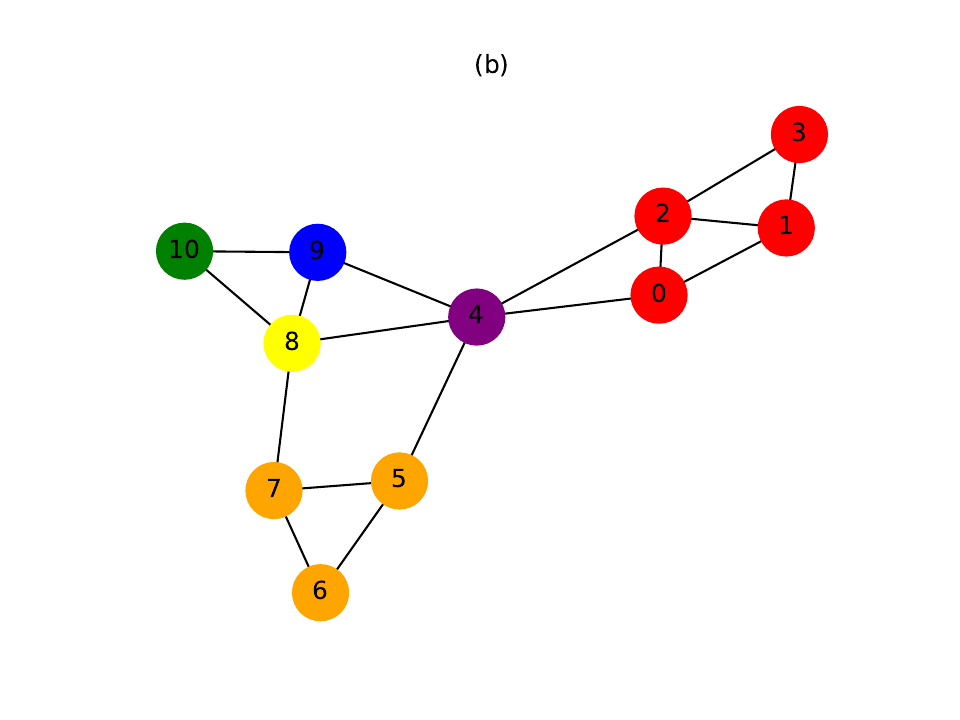} \quad
\caption{The partial destruction phase}
      \label{fig:sub1b}
    \end{subfigure}\quad
        \centering
    \begin{subfigure}[b]{.4\textwidth}
      \centering
    \includegraphics[scale=.32]{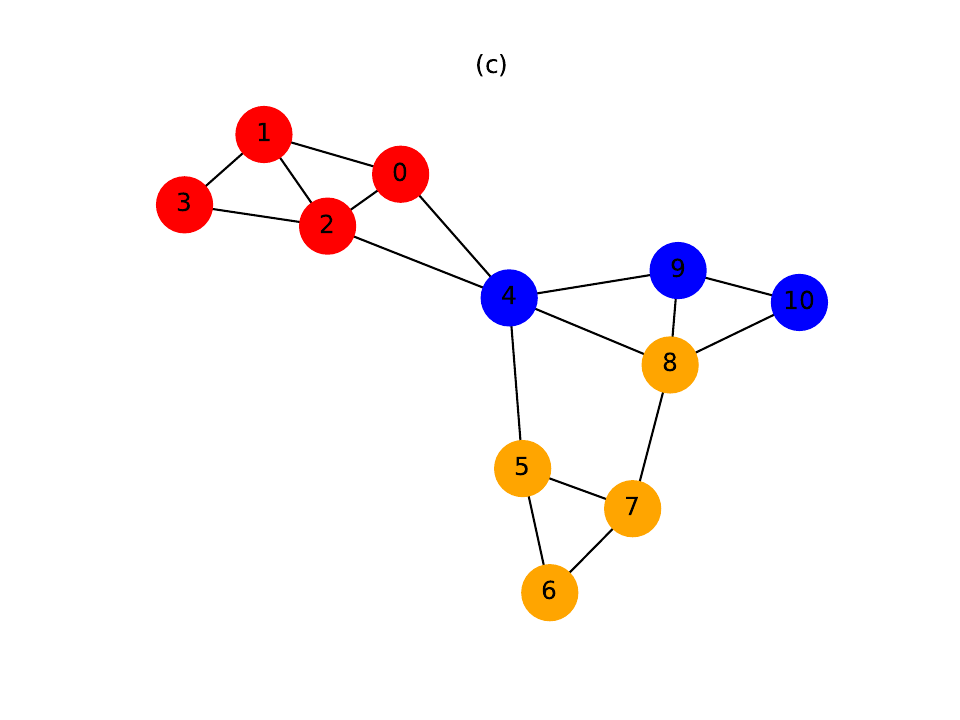}\quad
\caption{Random neighbor community technique}
      \label{fig:sub1b}
    \end{subfigure}\quad
        \begin{subfigure}[b]{.4\textwidth}
      \centering
    \includegraphics[scale=.32]{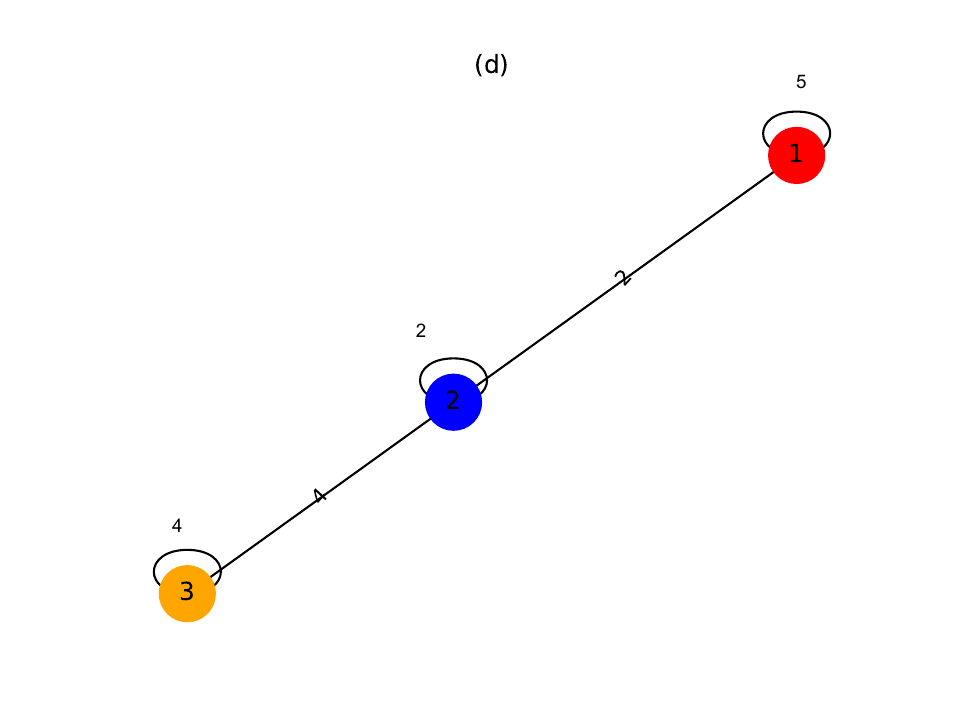}
\caption{The condense function}
      \label{fig:sub1b}
    \end{subfigure}\quad
         \begin{subfigure}[b]{.4\textwidth}
      \centering
     \includegraphics[scale=.32]{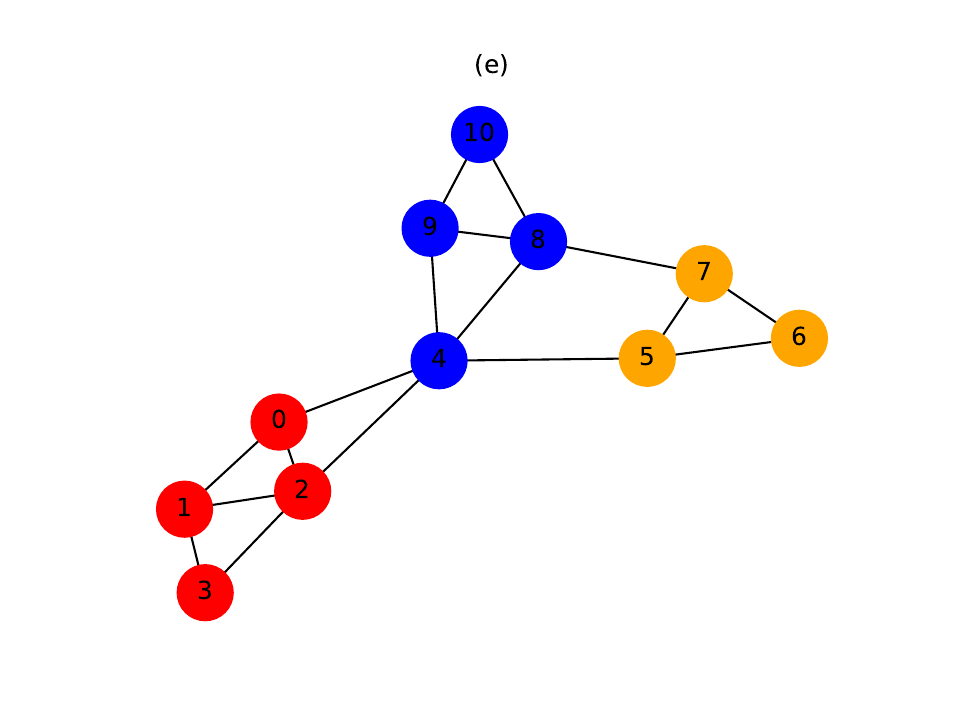}
\caption{The final solution}
      \label{fig:sub1b}
    \end{subfigure}\quad
\caption{Examples of FLMIG procedures}
\label{fig:flmig}
  \end{minipage}\\[1em]
\end{figure}

Algorithm \ref{algo1} offers a comprehensive examination of the FLMIG's approach for solving the community detection problems. The algorithm functions in the following manner: it initiates with the \textcolor{blue}{Generate Initial Solution} function to generate an initial solution. Subsequent iterations encompass multiple stages until a specific termination criterion is achieved. At first, \textcolor{blue}{the Destruction Solution function}  (step 4) disturbs a portion of the existing solution, usually by merging nodes into a single community. Next, the reconstruction phase (step 5) occurs, during which individual communities are merged into a random neighboring community using the \textcolor{blue}{Reconstruction Heuristic algorithm} . The prune Louvain algorithm\cite{bib24} is used to achieve fast convergence towards a feasible solution. The procedure is finalized by the \textbf{Select Next Solution} function, which determines if the actual constructed solution should be retained for the next iteration. The different phases of the FLMIG algorithm are illustrated in  Fig. \ref{fig:flmig}.  
\begin{algorithm}
\hspace*{\algorithmicindent} \textbf{Input}: Graph $G(V,E)$ \\
\hspace*{\algorithmicindent} \textbf{Output}: $S = \{C_1, C_2, C_3, \ldots, C_k\}$
\caption{Fast local move iterated greedy algorithm}\label{algo1}
\begin{algorithmic}[1]
\State $S^{*}$= \textbf{Generate Initial Solution} ($G$)
\State $S_{best} \gets S^{*}$
\Repeat
\State $S^{'}=$ \textbf{Destruction solution} $(S^{*})$
\State $S^{*'}=$ \textbf{Reconstruction Heuristic}   $(S^{'} )$
\State $ S^{*}=$ \textbf{Select Next Solution}    $(S^{*'},S^{*})$ 
\If{$Q(S^{*})$ $>$ $Q(S_{best})$ }
    \State $S_{best} \gets S^{*'}$
\EndIf
\Until Termination condition not satisfied
\end{algorithmic}
\end{algorithm}

\subsection{ \textcolor{blue}{Generate Initial Solution} Algorithm } \label{subsec1}
Similar to previous meta-heuristic methods, our Fast Local Move Iterated Greedy (FLMIG) algorithm begins by quickly generating an initial solution using the Generate Initial Solution function, as described in Algorithm \ref{algo2}. This function works by choosing a node $v$ from the candidate list $D$ and assigning it to a community $C_{i}$. It proceeds with this procedure, iterating through the list of candidates until it is depleted. After assigning each vertex to its corresponding community, \textcolor{blue}{as defined in steps 1 and 6 of Algorithm} \ref{algo2}, a local node movement is performed. The movement is made easier by the \textcolor{blue}{Fast Local Move} function, which is explained in depth in Algorithm \ref{algo3}. This function is used to provide an initial solution, as mentioned in step 8.

\begin{algorithm}
\hspace*{\algorithmicindent} \textbf{Input}: Graph $G(V,E)$ \\
\hspace*{\algorithmicindent} \textbf{Output}: $S=$\{$\mathcal{C}_{1}$,$\mathcal{C}_{2}$,$\mathcal{C}_{3}$,......,$\mathcal{C}_{k}$\}  
\caption{ \textcolor{blue}{Generate Initial Solution}}\label{algo2}
\begin{algorithmic}[1]
\State $S \gets \phi$
\State $CL \leftarrow V $
\While {$CL \not= \phi$} 
\State $v \gets SelectRandom(CL)$
\State $C_{i} $ $\leftarrow C_{i}  \cup  \{v\}$ 
\State $CL \gets CL $\textbackslash$ \{v\}$
\State $ S \gets S\cup \{C_{i}\} $
\EndWhile 
\State $S^{*} = FastLocalMove(S, G)$ \Comment{ See algorithm \ref{algo3}}
\State \textbf{Return}\ $S^{*}$
\end{algorithmic}
\end{algorithm}

\subsection{FastLocalMove Algorithm\label{subsec2}}
We have integrated the most recent improvements of the Louvain algorithm into the Iterated Greedy (IG) framework, which effectively resolves the constraints identified in earlier versions of IG\cite{bib21}\cite{bib22}. Usually, IG algorithms are improved by incorporating a Local Search (LS) metaheuristic. The main purpose of LS is to significantly enhance the search process by thoroughly examining the surrounding area of the solution for any possible enhancements, and only making changes to the nodes if they result in an increase in the modularity value.

LS evaluates all nodes in each iteration to compute the modularity change, $\Delta Q$. However, this frequently results in just a restricted number of nodes changing their community affiliations, leading to unnecessary operations and, thus, a delayed convergence. This is particularly troublesome in large-scale networks due to the increasing computational requirements. In order to address these problems, we suggest incorporating a Fast Local Move (FLM) heuristic into the IG algorithm\cite{bib24}. The objective of this adaption is to decrease the computational burden while maintaining solution quality and improving the search intensification. The FLM heuristic focuses on nodes that are more likely to shift between communities. It calculates changes in modularity based on these specific nodes, as described in Algorithm \ref{algo3}.

In the FLM technique, all nodes are initially placed in the queue $L{Q}$ (step 1). The approach involves illiterately removing a node $v$ from $L{Q}$ (step 3) and evaluating if it should be reassigned to a different community $C^{'}$ that provides the greatest increase in modularity. In addition, FLM defines $N$ as the set of neighbors of $v$ that are not in $C^{'}$, guaranteeing that they are taken into account in later iterations (steps 3 to 10).

In order to measure the increase in the modularity, denoted as $\Delta Q$, we employ vertex move (VM) operations as described in reference \cite{bib24}. Virtual machine (VM) entails relocating a node from its existing community to a neighboring one, resulting in a substantial reduction in computational time while preserving the algorithm's optimization process integrity. 

The modularity gain $\Delta Q$ of VM $(v,C,C^{'})$ can be formally defined:\\
\begin{equation}
    \Delta Q = e_{v,c^{'}} + \frac{d_v}{2m^2}(d_{tot}C^{'}) \label{eq3}
\end{equation}\\
Where $e_{v, c^{'}}$ is the number of edges between the community $v$ and $C^{'}$ and $d_{tot} \ C^{'}$ is the total degree of nodes in community $C^{'}$.   
      
\begin{algorithm}
\hspace*{\algorithmicindent} \textbf{Input}: Graph $G(V,E)$ \\
\hspace*{\algorithmicindent} \textbf{Output}: $S = \{C_1, C_2, C_3, \ldots, C_k\}$  
\caption{ Fast Local Move Algorithm. \label{algo3}}
\begin{algorithmic}[1]
\State $L_{Q}\gets Queue(V(G))$
\While {$L_{Q} \not= \phi$}
\State $v \gets L_{Q}.Remove()$
\State $\Delta Q = Max_{C^{'}\in S}(Q(C':\gets C{'}\cup \{v\}))$ 
\If  {$\Delta Q > 0$}
	    \State $C{'}\gets ArgMax_{C'\in S}Q(C{'}:\gets C{'}\cup \{v\})$
	    \State $C{'}\gets C{'}\cup \{v\}$
	    \State $C \gets C $\textbackslash$ \{v\}$
	    \State $N \{u \mid (u,v)\in E(G),u \not\in C'\}$
	    \State $L_{Q} = L_{Q}.enqueue(N-L_{Q})$
\EndIf
\EndWhile 
\State \textbf{Return}\ $S$
\end{algorithmic}
\end{algorithm}

\subsection{\textcolor{blue}{Destruction solution Algorithm}\label{subsec3}}
The destruction step in our technique is essential for perturbing the existing solution in order to generate a segmented version. The magnitude of this disturbance is determined by the parameter $\beta$. More precisely, a subset of nodes, determined by multiplying $\beta$ with $n$, is randomly selected from their current communities and moved to separate communities consisting of only one node. These nodes will be reassigned to different communities during the reconstruction phase. It should be emphasized that $\beta$ has a substantial impact on two crucial search mechanisms: intensification and diversification. However, in current versions of the Iterated Greedy (IG) algorithm, the optimal balance between these two methods has not been achieved. As a result, a lower value of $\beta$ tends to limit down the search too much, whereas a higher value of $\beta$ results in more random solutions during reconstruction. The inherent unpredictability can present difficulties in effectively reaching optimal solutions within a certain time range.
\begin{algorithm}
\hspace*{\algorithmicindent} \textbf{Input}: Graph $G(V,E),solution\ S $ \\
\hspace*{\algorithmicindent} \textbf{Output}: $ partial \ solution \ S, L_{n}$ 
\caption{Destruction solution}\label{algo4}
\begin{algorithmic}[1]
\State $L_{n}\gets \phi$
\State $L_{r} \gets V$   \Comment{put list of vertices in random order}
\For {$i=1$  to $\beta$}
\State $v=L_{r}.Remove()$
\State $L_{n}.add(v)$
\State $C \gets C $\textbackslash$ \{v\}$
\State $C_{i} $ $\leftarrow C_{i}  \cup  \{v\}$
\State $ S \gets S\cup \{C_{i}\} $
\EndFor 
\State \textbf{Return}\ $S ,$ \ $L_{n}$
\end{algorithmic}
\end{algorithm}
\subsection{ \textcolor{blue}{Reconstruction Heuristic} Algorithm\label{subsec4}}
The Fast Local Move Iterated Greedy (FLMIG) method effectively creates a strong equilibrium between the two essential search mechanisms: intensification and diversification, as stated earlier. This balance increases the probability of FLMIG producing community partitions of excellent quality.

After carrying out the \textcolor{blue}{Destruction solution procedure}, the subsequent phase is dedicated to the reintegration of singleton communities into their respective communities. The \textcolor{blue}{Reconstruction heuristic} function does this by doing two main acts. At first, $\beta$ nodes from the list $L_{n}$ are randomly and separately moved from their single communities to neighboring communities. The selection of this relocation is determined by employing the $SelectRandomCommunity$ function. This function differs from standard greedy approaches by allowing nodes to be moved into any community that improves the modularity function, even if the community is randomly chosen. The likelihood of selecting a specific community is directly related to the extent to which moving a node to that community would result in an improvement in modularity. This strategy facilitates a more varied and subtle process of combining communities.

The function $SelectRandomCommunity$ is formally described as follow:

\begin{equation}
 Pr(C^{'} = C) =
\begin{cases}
exp(\frac{1}{\epsilon} \Delta Q(v \longrightarrow C)) & if \Delta Q(v \longrightarrow C) \geq 0 \\
0 & \text otherwise
\end{cases}
\label{eq4}
\end{equation}
\\
\\
After the initial phase, the improved prune Louvain algorithm starts using the solution obtained up to that point. As mentioned above, the prune Louvain could produce disconnected communities. Thus, a new technique is adopted to address this shortcoming. The operation is structured as follows: The obtained solution is refined. Subsequently, nodes belonging to the same partitions are combined to create a condensed graph G. Afterwards, the quick local move heuristic is used to determine the community structure in the new graph G'. This approach is iteratively performed until no further enhancements in the modularity value are observed (as described in steps 12 and 19 in Algorithm  \ref{algo5}).
The Refinement Communities procedure is dedicated to refining the partition obtained. The function does the following steps iteratively until the partition is \textcolor{blue}{well-connected}:
In the first step, define the disconnected partitions. The second step is to reassign each node in the same disconnected component into a new community. The third step is to move these nodes locally to maximize the gain in modularity.   Implementing a random neighbor selection technique not only expands the range of search but also increases the likelihood of finding a wide variety of answers. By including the prune Louvain algorithm, these techniques greatly accelerate the convergence of the FLMIG algorithm towards the global optimum. The FLMIG algorithm derives advantages from these approaches by simultaneously enhancing intensification and diversification. This simultaneous impact enables a harmonious examination of the range of possibilities, successfully combining both extensive inquiry and precise experience-driven search strategies. Moreover, these solutions enhance the overall stability and efficiency of the FLMIG algorithm.

\begin{algorithm}
\hspace*{\algorithmicindent} \textbf{Input}: Graph $G(V,E), Incumbent\ solution\ S,L_{n}$ \\
\hspace*{\algorithmicindent} \textbf{Output}: $S$ 
\caption{ Reconstruction Heuristic}\label{algo5}
\begin{algorithmic}[1]
\While{$L_{n}\not= \phi$}
\State $v \gets SelectRandom(L_{n})$
\State $L_{n} \gets L_{n} $\textbackslash$ \{v\}$
\State $\Delta Q \gets  Md_{\ C^{'}\in S} (Q(C^{'} :\gets  C^{'}  \cup  \{v\}))$
\If  {${\Delta  Q> 0 }$}
	    \State $C^{'} \gets$ $SelectRandomCommunity(C^{'})$
	    \State $C^{'} \gets C^{'} \cup  \{v\}$
	    \State $C \gets C $\textbackslash$ \{v\}$
\EndIf
\EndWhile
\State $Improvment = True $
\While{$Improvment = True$}
\State $Improvment = False $
\State $S \gets Refinement Communites(P)$     \Comment{ See algorithm \ref{algo6}}
\State $G^{'} \gets Condence(G, S)$
\State $ S^{'} = Fastlocalmove(G^{'},S)$ \Comment{ See algorithm \ref{algo3}}
\State $Q_{new} \gets Q_{S^{'}} $
\If  {${Q_{new} > Q(S^{'}) }$}
	    \State $Improvment = True $
\EndIf
\EndWhile
\State \textbf{Return}\ $S^{'}$
\end{algorithmic}
\end{algorithm}

\begin{algorithm}
\hspace*{\algorithmicindent} \textbf{Input}:  $Unvalidated$ $Partitions$ $\mathcal{P}$ \\
\hspace*{\algorithmicindent} \textbf{Output}: $S = \{C_1, C_2, C_3, \ldots, C_k\}$   
\caption{ Refinement Communities}\label{algo6}
\begin{algorithmic}[1]
\State $ Refinment = False $
\While{$ Refinment = False$}
\State $D \gets DisconnectedPartition(P)$
\State $C_{new} \gets ReassignedPartition(D)$  
\State $S \gets LocalMoveNode(G, C_{new})$
\If   {$ S = Connected$  }
         \State $ Refinement = True$
\EndIf
\EndWhile
\State \textbf{Return}\ $S$          
\end{algorithmic}
\end{algorithm}

\subsection{\textcolor{blue}{Select Next Solution Algorithm} \label{subsec5}}
The ultimate function in our approach assesses whether the current solution should be maintained as the major solution for the next iteration. The widely used 'Replace if Better' (RB) criterion, noted for its simple implementation, frequently faces problems with stagnation caused by insufficient diversification. In order to increase the variety of options in the search space, we employ simulated annealing as our acceptance criterion. This enables us to consider solutions with a certain probability.

To be more precise, a new solution $S^{'}$ is considered acceptable if it exhibits greater modularity than the current solution $S$. If not, the acceptance of $S^{'}$ is still evaluated, but with a probability defined by the expression $exp\left(\frac{Q(S')-Q(S)}{T}\right)$, where $T$ represents a control parameter referred to as temperature. In accordance with the rules established by Stützle\cite{bib46}, we begin by locally optimizing the initial solution $S^0$ in order to generate $S$. The initial temperature, denoted as $T$, is thereafter assigned a value equal to 0.025 multiplied by the modularity of $S^{*}$. An acceptable solution is defined as one that is up to 2.5\% less effective than the existing solution, with a probability of $1/e$. The temperature $T$ is progressively reduced in successive repetitions, undergoing a multiplication by 0.9 each time, following the geometric cooling schedule. This technique successfully achieves a harmonious combination of exploring and intensifying the search, hence improving the overall performance and diversity of the researched options.

\subsection{Complexity analysis}\label{subsec6}
In this paper, we provide a comprehensive examination of the computational complexity of the Fast Local Move Iterated Greedy (FLMIG) algorithm. The time complexity of the \textcolor{blue}{Generate Initial Solution} function is assessed as $O(n \times (n-1)/2)$, where $n$ represents the input size. Regarding the \textcolor{blue}{Destruction Solution} function, the computing cost of eliminating $\beta$ nodes from the existing solution is measured as $O(\beta)$. During the reconstruction phase, the cost of selecting a random neighbor community remains at a complexity of $O(\beta)$. On the other hand, performing the prune Louvain method has a complexity of $O(n \log(n))$. Therefore, the total computing cost for the \textcolor{blue}{Reconstruction Heuristic} function is $O(\beta + n\log(n))$.

Assuming that the $\mathit{FastLocalMove}$ function produces an improvement after $t$ iterations, its worst-case complexity can be estimated as $O(2t \times m)$. After taking into account all of these factors, the overall computational complexity of the FLMIG method can be estimated as $O(n \times (n-1)/2 + T \times (1+\beta+ n\log(n)+ 2m\times t))$. However, this can be simplified and closely approximated as $O(n^2)$. This analysis offers a comprehensive comprehension of the computational requirements and effectiveness of the FLMIG algorithm in different contexts.

\section{Experimental Evaluation}\label{sec4}
In this section, we provide the results obtained using the FLIMG method and offer a comprehensive evaluation of its performance through computational tests. The Python programming language was used to implement all algorithms, including FLIMG, with the utilization of the NetworkX module. The tests were conducted on a system running on an Intel Core i5-13600KF CPU clocked at 5.1 GHz and equipped with 16 GB of RAM, operating on Ubuntu 22.04 LTS.

In order to prove the efficacy and efficiency of the proposed approach FLIMG algorithm, we compare it with four well-known approaches in the literature, namely: DBAT-M, IG, ICG, and CC-GA. Furthermore, five heuristic methods were used for the purpose of conducting comparative analysis. The algorithms mentioned include FN\cite{bib31}, Louvain\cite{bib11}, Louvain Prune\cite{bib24}, \textcolor{blue}{Leiden \cite{bib12}} and Label Propagation Algorithm (LPA)\cite{bib58}. The wide array of algorithms provide a strong foundation for evaluating the efficacy and efficiency of the FLIMG approach in many computational contexts.
\subsection{ Problem instances}\label{subsec1}
The evaluation test cases are classified into two main categories: real-world networks and synthetically generated networks. These benchmarks, including the GN benchmark\cite{bib49} and the LFR benchmark\cite{bib57}, are well respected in the field for assessing the performance of network algorithms.
\subsubsection{Real-world networks}\label{subsubsec1}
 The dataset used for testing comprises 12 real-world networks, frequently utilized in various algorithmic evaluations. These networks are categorized by size:

    \begin{enumerate}
    \item Small networks include:
     \begin{itemize}
     \item Zachary's Karate Club network\cite{bib47}
     \item Dolphins social network\cite{bib48}
     \item American college football network\cite{bib49}
     \item Co-purchased political Books network (polbooks)\cite{bib50}
     \item Les Miserables network (lesmis)\cite{bib51}
     \item Word adjacencies network (Adjnoun)\cite{bib50}
     \item Jazz musicians network (Jazz)\cite{bib52}
     \item C.elegans Metabolic network (Metabolic)\cite{bib53}
     \end{itemize}

   \item Medium-sized networks include:
       \begin{itemize}
       \item Scientist co-authorship network (Netscience)\cite{bib50}
       \item PGP social network (Pretty-good privacy)\cite{bib54}
       \item Internet network (As-22july06)\cite{bib55}
       \end{itemize}

   \item Large networks include:
     \begin{itemize}
     \item DBLP co-authorship network\cite{bib56}
      \item Amazon website network\cite{bib56}
     \end{itemize}
    \end{enumerate}

Among these, four small networks have a known community structure. Detailed characteristics of these networks are summarized in Table\ref{tab1}.

\begin{table*}[t]
\begin{center}
\setlength{\arrayrulewidth}{0.5mm}
\caption{Overview information of real-world networks}\label{tab1}    

\begin{tabular}{  p{2cm}  p{2cm}  p{2cm}  p{2cm}  p{2cm}  p{2cm} } 
\toprule
Networks & $\mid V \mid$ & $\mid E \mid$ & Avg degree & Avg CC & Ground-truths \\ [1ex] 
\midrule
Karate & 34 & 78 & 4.588 & 0.588 & 2 \\
Dolphins & 62 & 160 & 5.129 & 0.2859 & 2 \\
Polbooks & 105 & 441 &  8.4 & 0.4875 & 3 \\
Football & 115 & 613 & 10.661 & 0.4033  & 12 \\
lesmis  & 77   & 254 & 6.597 & 0.736  & Unknown \\
Adjnoun & 112 & 425 & 7.589 & 0.190  & Unknown \\
Jazz & 198 & 2742 & 27.70 & 0.633  & Unknown \\
Metabolic & 453 & 2040 & 8.940 & 0.655 & Unknown \\
NetScience & 1589 & 2742 & 3.451 & 0.6377 & Unknown \\
PGP & 10680 & 24316 & 4.5 & 0.2659 & Unknown \\
As-22july06 & 22963  & 48436 & 4.21 & 0.2304 & Unknown \\
com-DBLP & 317080 & 1049866 & 5.530 & 0.732 & Unknown \\
com-Amazon & 334863 & 925872 & 6.622 & 0.430 & Unknown \\
\bottomrule
\end{tabular}
\end{center}
\end{table*}

\subsubsection{Synthetic networks}\label{subsubsec2}
The FLMIG method was evaluated on two types of synthetic networks with specified community structures: the Girvan-Newman (GN) networks\cite{bib49} and the Lancichinetti-Fortunato-Radicchi (LFR) networks\cite{bib57}.

Firstly, the GN networks are composed of 128 nodes that are organized into four communities, with each community containing 32 nodes. Ten diverse networks were generated to evaluate distinct community architectures, controlled by the mixing parameter $u$. Within these networks, every node establishes connections with a proportion (1 - $u$) of its connections within its own community, while allocating a fraction $u$ of its connections to nodes outside its community.

Furthermore, the LFR networks, formulated by Lancichinetti, Fortunato, and Radicchi, provide a more accurate representation by including the diversity in both the connectivity and size of communities. These networks similarly exhibit a power-law distribution controlled by parameters $\theta_{1}$ and $\theta_{1}$. Two sets of LFR benchmarks were generated, each consisting of fourteen distinct networks. The construction of these networks required adjusting the mixing parameter $u$ in order to achieve a proper balance between the exterior and internal degree proportions of nodes. The specifications for these artificial networks, such as the number of nodes ($N$), average node degree ($K$), maximum node degree ($MaxK$), minimum community size ($Minc$), and maximum community size ($Maxc$), are provided in Table \ref{tab2}.

\begin{table*}[h!] 
\addtolength{\tabcolsep}{-6pt}
\caption{Parameter settings of LFR benchmark networks} \label{tab2}
\begin{center}  
\begin{tabular}{ p{2cm}  p{2cm}  p{2cm} p{2cm}}
\toprule
Networks & LFR(A) & LFR(B) & GN \\ [1ex] 
\midrule
 N & \textcolor{blue}{$10^{3}$} & \textcolor{blue}{$10^{3}$,$10^{4}$,$10^{5}$} & 128 \\ 
 K & 15 &  15   & 16\\
 MaxK & 100 &  50 & 16 \\
 Minc & 20 &  10 & 32 \\
 Maxc  & 100 & 50 & 32  \\
 $\mu$ & \textcolor{blue}{0.2 - 0.8} & \textcolor{blue}{0.2 - 0.8}& 0.1 - 0.5 \\
 $\theta_{1}$& 2 & 2 & 2 \\
 $\theta_{2}$& 1 & 1 & 1 \\ [2ex]
 \bottomrule
 \end{tabular}
 \end{center}
\end{table*}

\subsection{Evaluation criterion}\label{subsec2}
To assess the efficiency of our proposed approach, we have employed widely used metrics that gauge the accuracy of detected network partitions. The measures used in this study are modularity ($Q$) and normalized mutual information (NMI) \cite{bib59}. The normalized mutual information (NMI) is especially valuable in situations where the precise divisions of the network have previously been determined. The NMI (Normalized Mutual Information) is a numerical measure that falls within the range of 0 to 1. A number closer to 1 implies a stronger similarity between the discovered partitions and the actual partitions. This metric is essential for comprehending the degree to which the algorithm's output corresponds with the established community structures within the networks.

\begin{equation}
NMI(A,B) = \frac{-2\sum_{i=1}^{M_{A}}\sum_{j=1}^{M_{B}} M_{ij} \log\left(\frac{n \cdot M_{ij}}{M_{i.}M_{.j}}\right)}{\sum_{i=1}^{M_{A}}M_{i.}\log\left(\frac{M_{i.}}{n}\right)+\sum_{j=1}^{M_{B}}M_{.j}\log\left(\frac{M_{.j}}{n}\right)}
\label{eq4}
\end{equation}
\large

\subsection{ Fine tuning algorithm }\label{subsec3}
The optimal performance of the FLMIG algorithm relies heavily on the effective tuning of parameters $\beta$ and $\epsilon$, \textcolor{blue}{the source code of the FLMIG algorithm is available at \href{https://github.com/salahinfo/FLMIG_algorithm.git}{https://github.com/salahinfo/FLMIG\_algorithm}}.

To enhance these values, we conducted comprehensive experiments on multiple actual networks. The approach was crucial in identifying the most favorable values for $\beta$, and we observed that the computational outcomes were constant across multiple networks. An illustrative example is the Lesmis network, where the impact of various $\beta$ values (ranging from 0.1 to 0.9) on both modularity and computational time was thoroughly examined. In order to assess its influence, we observed the average modularity for each $\beta$ configuration.

In addition, we have defined a termination condition, represented by the variable $r$, which indicates the maximum number of repetitions permitted without any enhancement in modularity. The value of variable $r$ was assigned according to the size of the networks: 100 for tiny networks, 50 for medium networks like Netscience, PGP, and As-22july06, and 10 for bigger networks such as DBLP and Amazon.

\begin{figure}[t]%
\centering
\includegraphics[scale=.5]{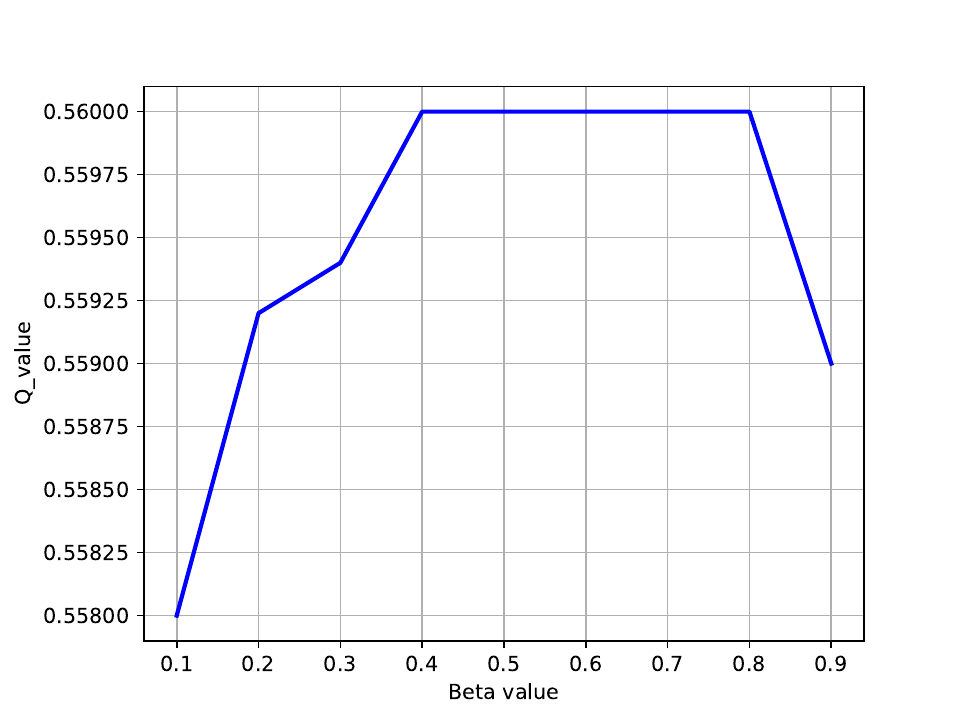}
\caption{ The impact of $\beta$ values on the modularity value.}
\label{fig2}
\end{figure}

\begin{figure}[h]%
\centering
\includegraphics[scale=.5]{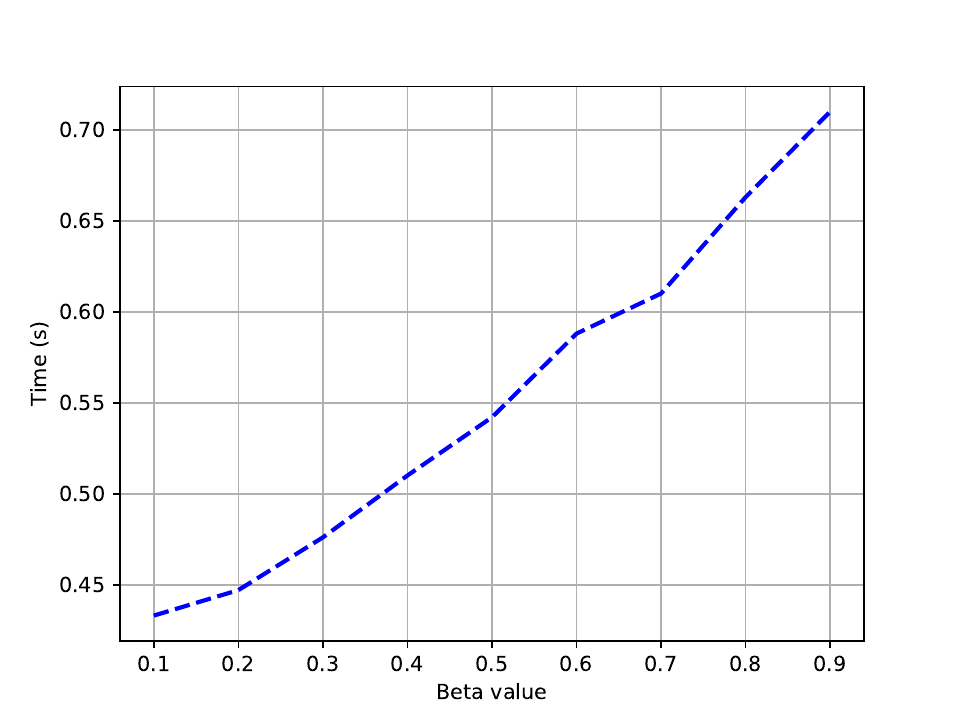}
\caption{ The impact of $\beta$ values on the computational time.}
\label{fig3}
\end{figure}

The impacts of different $\beta$ values can be examined in Figures Fig.  \ref{fig2} and Fig.  \ref{fig3}. Our experiments determined that setting the randomization parameter $\epsilon$ to 0.01 produced good results. The range of 0.01 to 0.1 was considered for $\epsilon$. The precise parameter configurations for each network may be found in Table \ref{tab3}, while the parameters employed for the comparative methods are presented in Table \ref{tab4}. The thorough examination of parameters had a crucial role in improving the performance of the FLMIG algorithm.

\begin{table}
\caption{Fine tuning of the FLMIG algorithm}\label{tab3}%
\begin{tabular}{ p{2cm}  p{4cm} }
\hline
\\[0.5pt]
Networks & Parameters and Values \\
\\[0.5pt]
\hline
\\[0.5pt]

Karate &  $\beta = 0.5$,    $nbr = 100$ \\
Dolphins &  $\beta = 0.7$,  $nbr = 100$ \\
Football &  $\beta = 0.5$,  $nbr = 200$ \\
Polbooks &  $\beta = 0.4$,  $nbr = 200$ \\
Adjnoun &  $\beta = 0.5$,   $nbr = 300$ \\
Lesmis &  $\beta = 0.4$,    $nbr = 200$ \\
Metabolic &  $\beta = 0.5$, $nbr = 300$ \\
Netscience &  $\beta = 0.5$, $nbr = 100$ \\
PGP &  $\beta = 0.5$,        $nbr = 100$ \\
As-06jully22 &  $\beta = 0.5$, $nbr = 100$ \\
com-Amazon &  $\beta = 0.4$,    $nbr = 20$ \\
com-dblp &  $\beta = 0.4$,     $nbr = 20$ \\

\hline
\end{tabular}   
\end{table}

\begin{table*}
\caption{Parameters setting of compablue algorithms}\label{tab4}
\begin{tabular}{p{2cm} p{2cm} p{3cm} p{5cm}}
\hline
Algorithm & Parameters & Value & Description \\
\hline
\textbf{IG} & $\beta$ & 0.5 & Destruction rate \\
            & $nbr$ & 300 & Number of iteration \\
            & $\alpha$ & 0.3 & Carousel rate \\
\textbf{ICG} & $\beta$ & 0.5 & Destruction rate \\
             & $nbr$ & 300 & Number of iteration \\
             & $P_{c}$ & 200 & Initial population \\
\textbf{CC-GA} & $P_{s}$ & 0.2 & Crossover rate \\
               & $P_{m}$ & 0.15 & Mutation rate \\
               & $\alpha$ & 0.02 & Mutation extension \\
               & $T_{c}$ & 0.1 & Elite reproduction of the existing population \\
               & $nbr$ & 200 & Number of iteration \\
\textbf{DBAT-M} & $R$ & 0.5 & Pulse rate \\
                & $A$ & 0.6 & Loudness \\
                & $D$ & number of nodes & Dimension of solution \\
                & $nbr$ & 100 & Number of iteration \\
\hline
\end{tabular}
\end{table*}

\section{ Results and Discussion} \label{sec5}

\subsection{ Convergence analysis} \label{subsec1}
\begin{figure}[H]
 \advance\leftskip-2cm
  \begin{minipage}{1.2\textwidth}
    \centering
    \begin{subfigure}[b]{.45\textwidth}
      \centering
    \includegraphics[scale=.45]{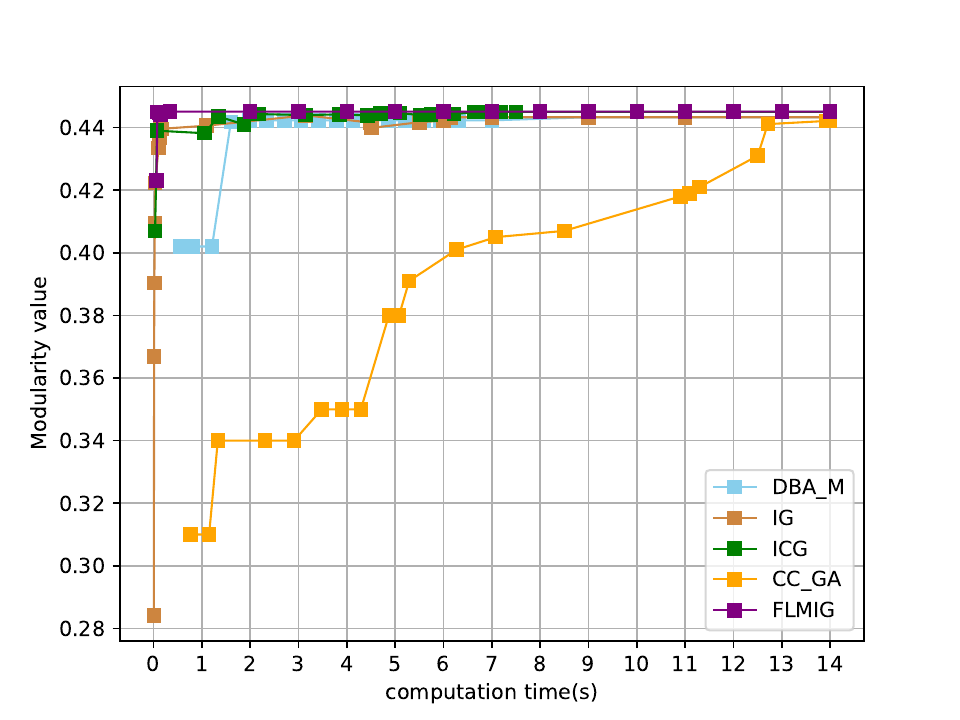} \quad
    \caption{Jazz network}
      \label{fig:sub1a}
    \end{subfigure}\quad
    \centering
    \begin{subfigure}[b]{.45\textwidth}
      \centering
    \includegraphics[scale=.45]{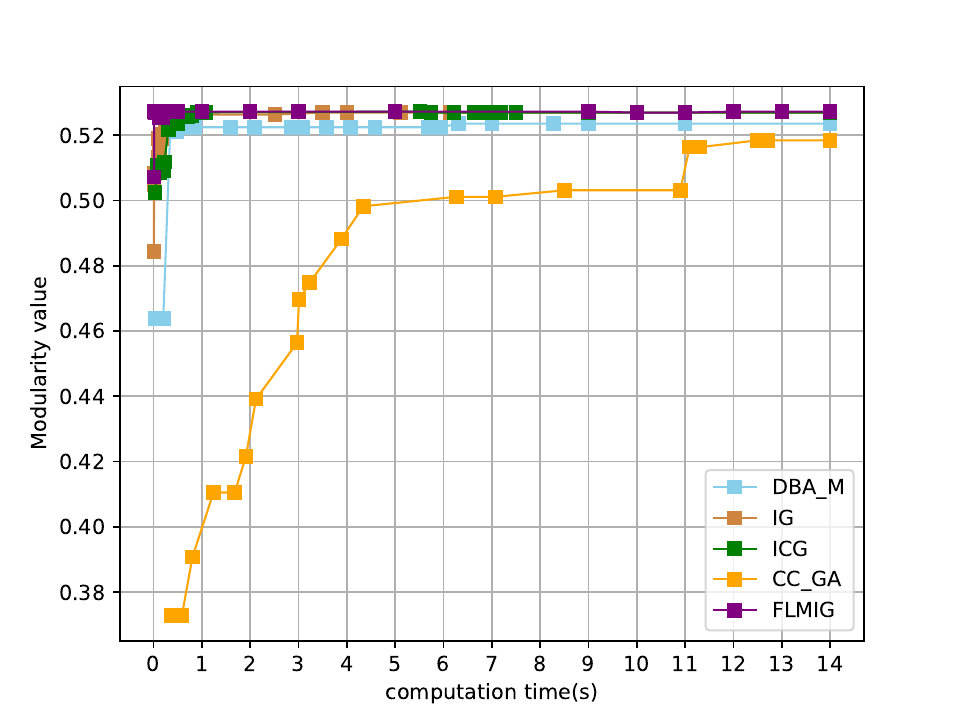} \quad
\caption{Polbooks network}
      \label{fig:sub1b}
    \end{subfigure}\quad
        \centering
    \begin{subfigure}[b]{.45\textwidth}
      \centering
    \includegraphics[scale=.45]{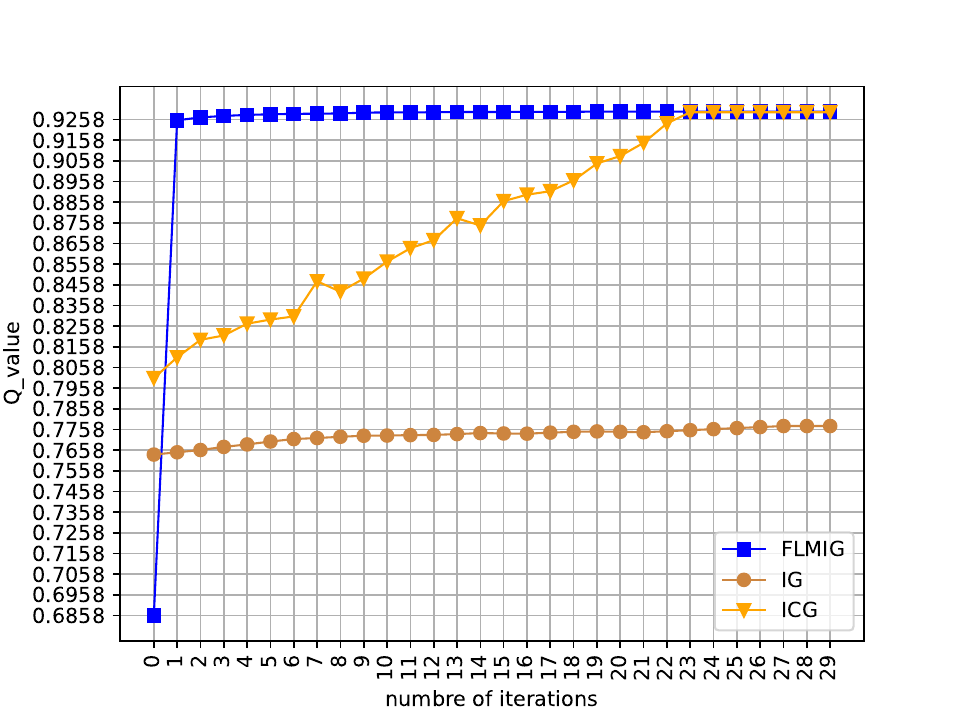} \quad
\caption{Amazon networks}
      \label{fig:sub1b}
    \end{subfigure}\quad
        \centering
    \begin{subfigure}[b]{.45\textwidth}
      \centering
    \includegraphics[scale=.45]{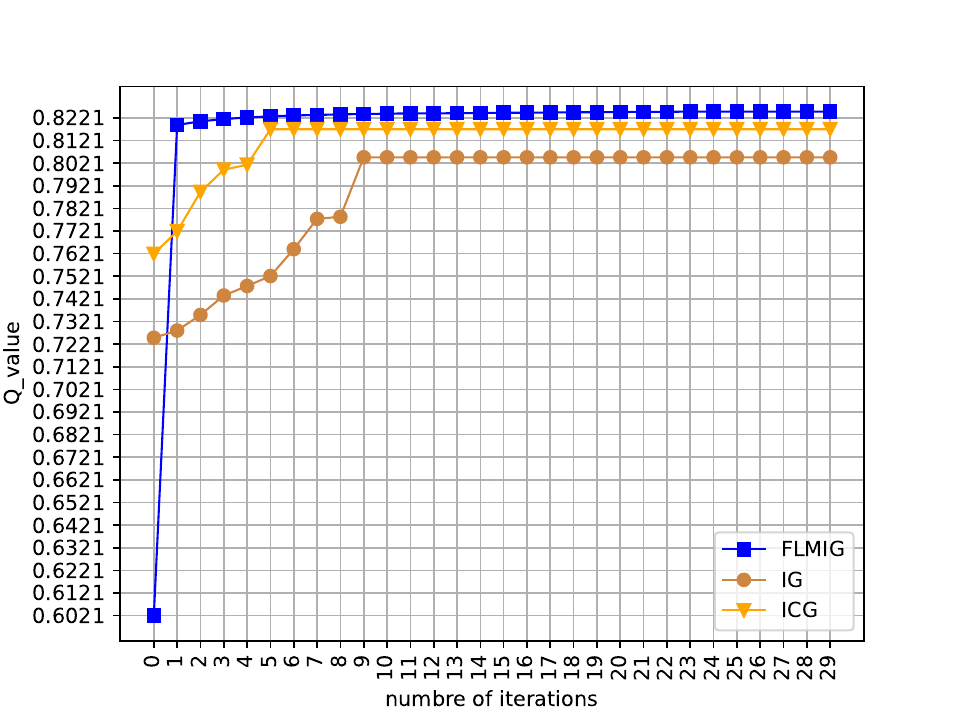} \quad
\caption{Dblp networks}
      \label{fig:sub1b}
    \end{subfigure}\quad
        \centering
\caption{Convergence analyse of FLMIG, IG, ICG, CC-GA and DBAT-M algorithms}
\label{fig4}
  \end{minipage}\\[1em]
\end{figure}

For each algorithm examined, our attention was on monitoring the growth of modularity values to comprehend the convergence characteristics of FLMIG in relation to other metaheuristics. In order to accomplish this, we chose two particular networks, namely Jazz and Polbooks, for an in-depth examination, as depicted in Fig. \ref{fig4} (a). and Fig.  \ref{fig4} (b). The x-axis in these pictures shows the computational time in seconds, while the y-axis corresponds to the modularity values.

The results of our study demonstrate that the FLMIG algorithm outperforms metaheuristics such as ICG, IG, DBAT-M, and CC-GA in terms of both the number of iterations and computational time required to obtain the global optimum. In the Polbooks network, optimality was achieved in a mere 0.014 seconds, but in the Jazz network, it required 0.08 seconds. FLMIG initially develops a solution with a significantly high modularity value and then iteratively improves this value to quickly attain the global optimum. An important characteristic of the FLMIG's strategy is the initial swift growth in modularity, which is subsequently reduced as a result of the simulated annealing (SA) acceptance criterion.

In addition, FLMIG showcases its capacity to effectively achieve the best possible outcome in extensive networks with minimal iterations, as depicted in Fig.  \ref{fig4} (c). and Fig.  \ref{fig4} (d). The performance of the algorithm demonstrates its resilience and efficiency across different network sizes.

\subsection{ Performance on real-world networks} \label{subsec2}
\begin{figure}[H]
 \advance\leftskip-2cm
  \begin{minipage}{1.2\textwidth}
    \centering
    \begin{subfigure}[b]{.45\textwidth}
      \centering
    \includegraphics[scale=.25]{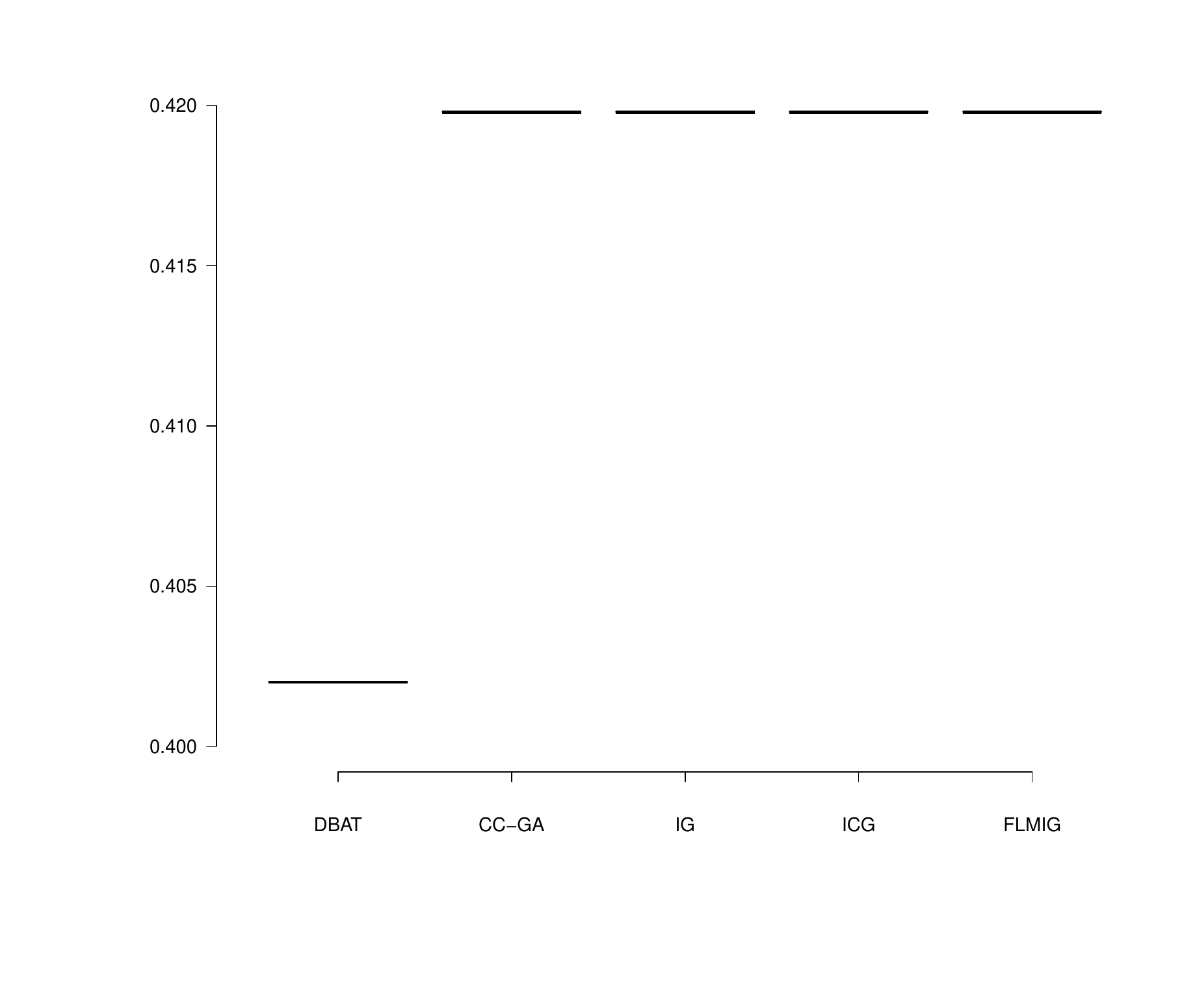} \quad
    \caption{Karate}
      \label{fig:sub1a}
    \end{subfigure}\quad
    \centering
    \begin{subfigure}[b]{.45\textwidth}
      \centering
    \includegraphics[scale=.25]{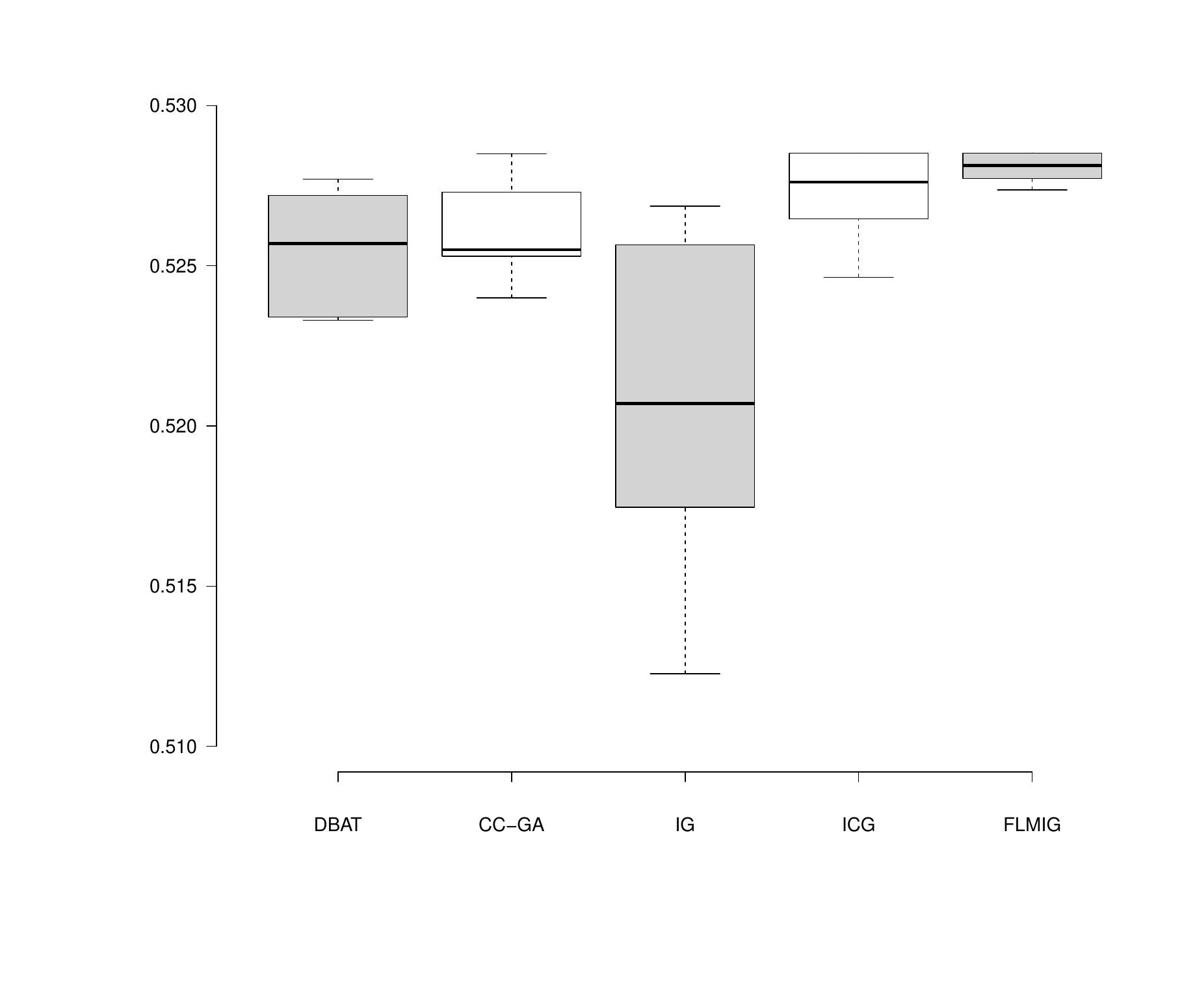} \quad
\caption{Dolphins}
      \label{fig:sub1b}
    \end{subfigure}\quad
        \centering
    \begin{subfigure}[b]{.45\textwidth}
      \centering
    \includegraphics[scale=.25]{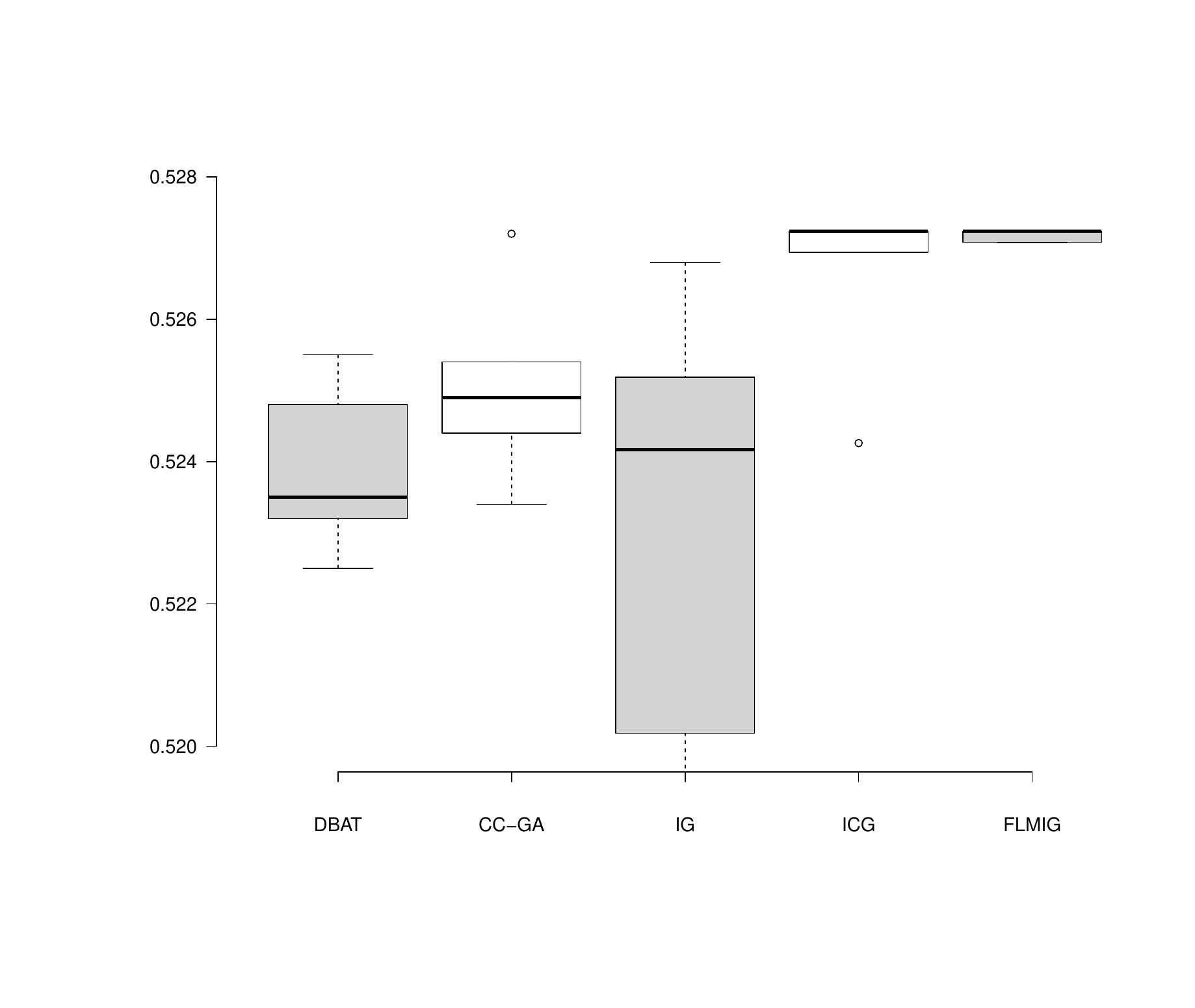} \quad
\caption{Polbooks}
      \label{fig:sub1b}
    \end{subfigure}\quad
        \centering
    \begin{subfigure}[b]{.45\textwidth}
      \centering
    \includegraphics[scale=.25]{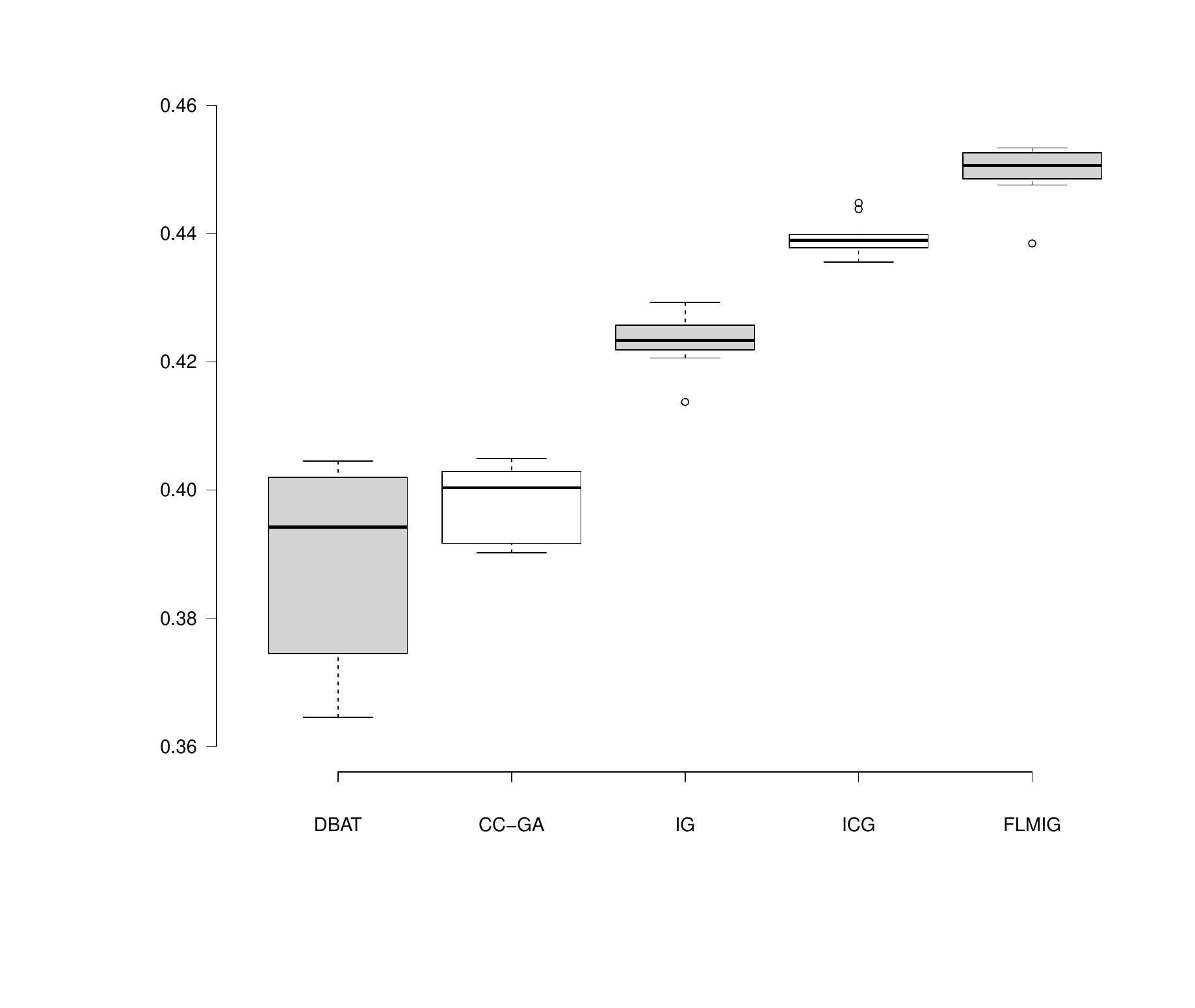} \quad
\caption{Metabolic}
      \label{fig:sub1b}
    \end{subfigure}\quad
        \centering
    \begin{subfigure}[b]{.45\textwidth}
      \centering
    \includegraphics[scale=.25]{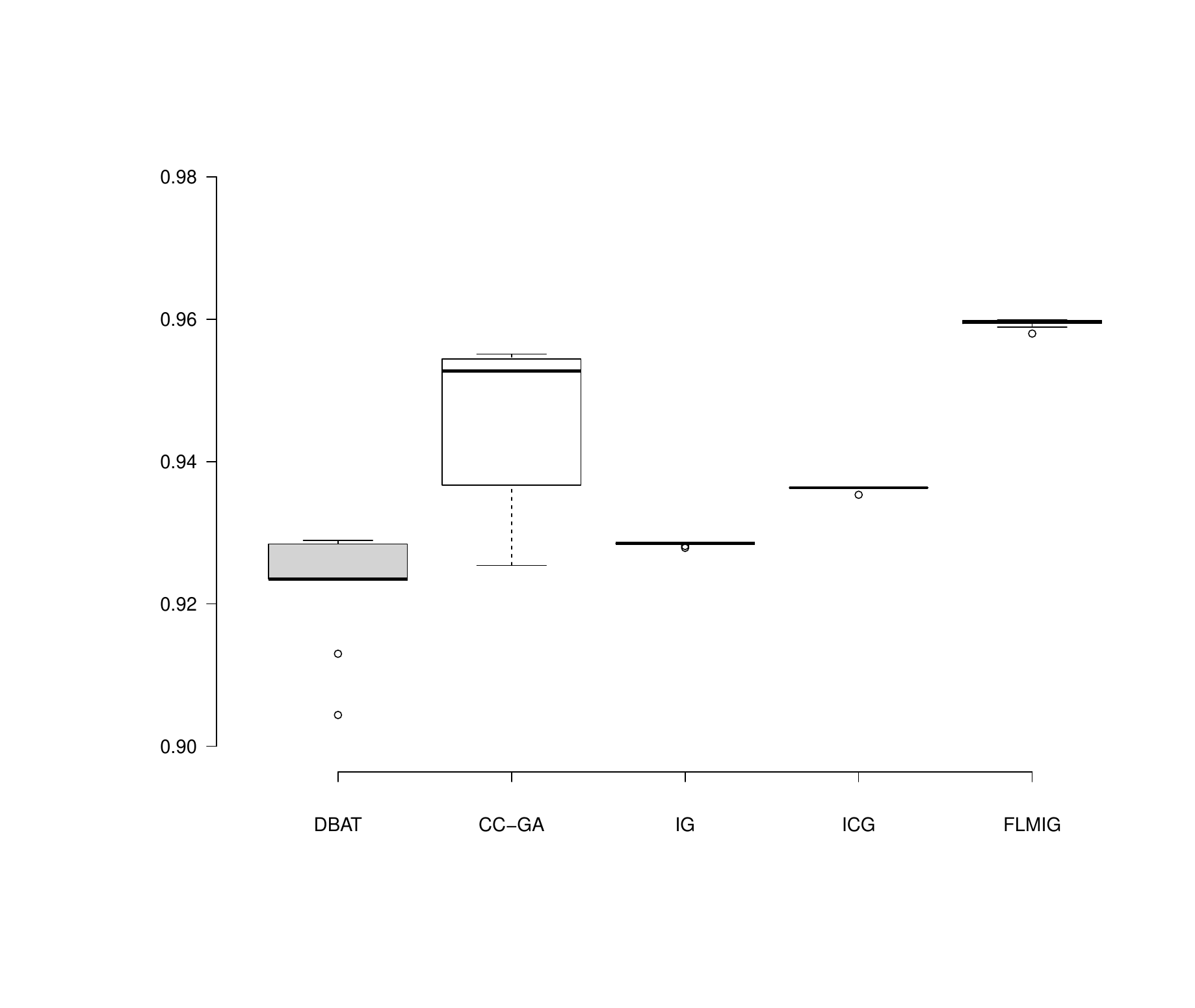} \quad
\caption{Netscience}
      \label{fig:sub1b}
    \end{subfigure}\quad
        \centering
    \begin{subfigure}[b]{.45\textwidth}
      \centering
    \includegraphics[scale=.25]{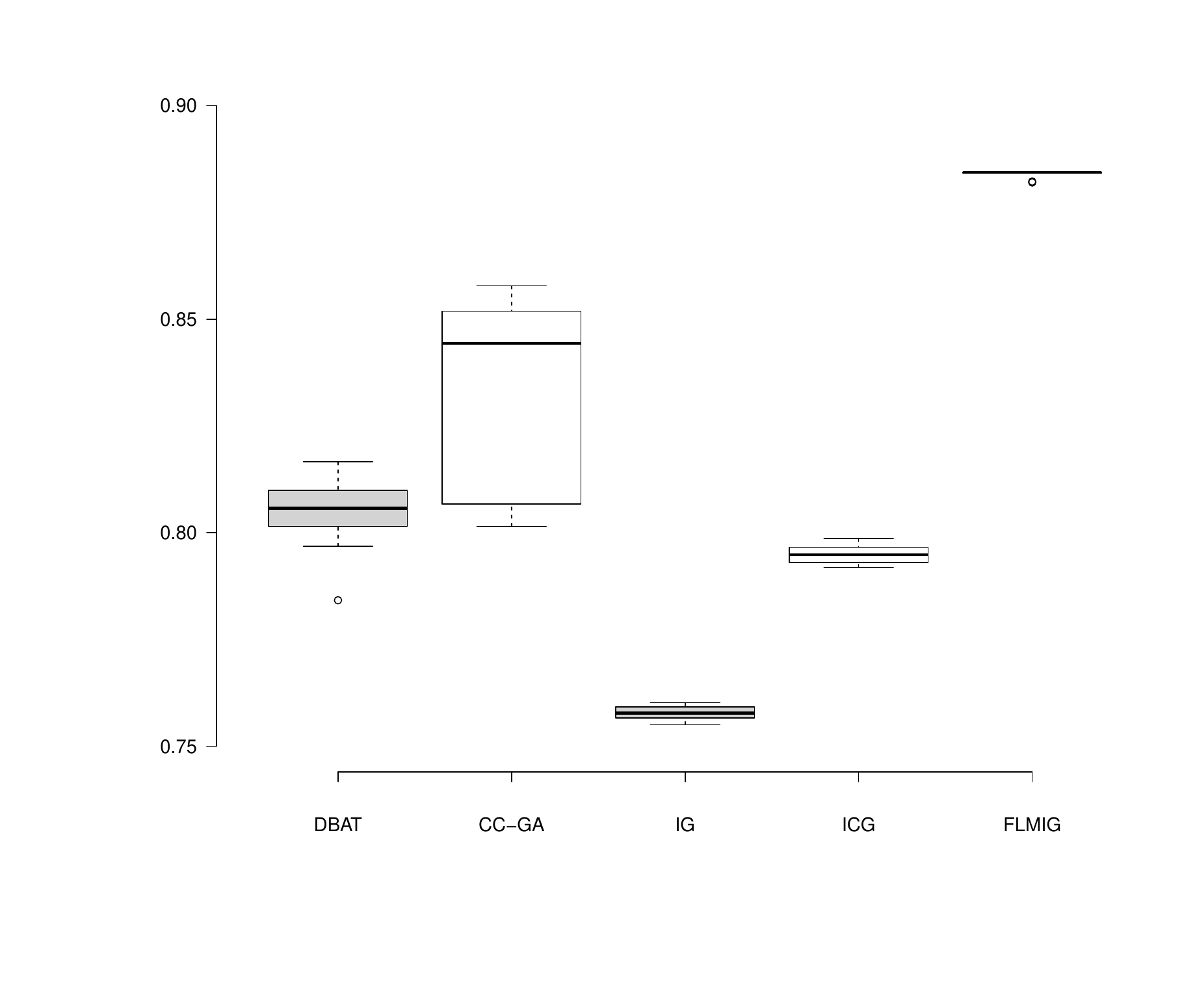} \quad
\caption{PGP}
      \label{fig:sub1b}
    \end{subfigure}\quad
      \caption{Experimental results of FLMIG and the compered algorithms on smaller and medium size real-world networks (1)}
\label{fig8}      
  \end{minipage}\\[1em]
  \end{figure}

\begin{figure}[H]
   \advance\leftskip-2cm
  \begin{minipage}{1.2\textwidth}
    \centering
    \begin{subfigure}[b]{.45\textwidth}
      \centering
    \includegraphics[scale=.25]{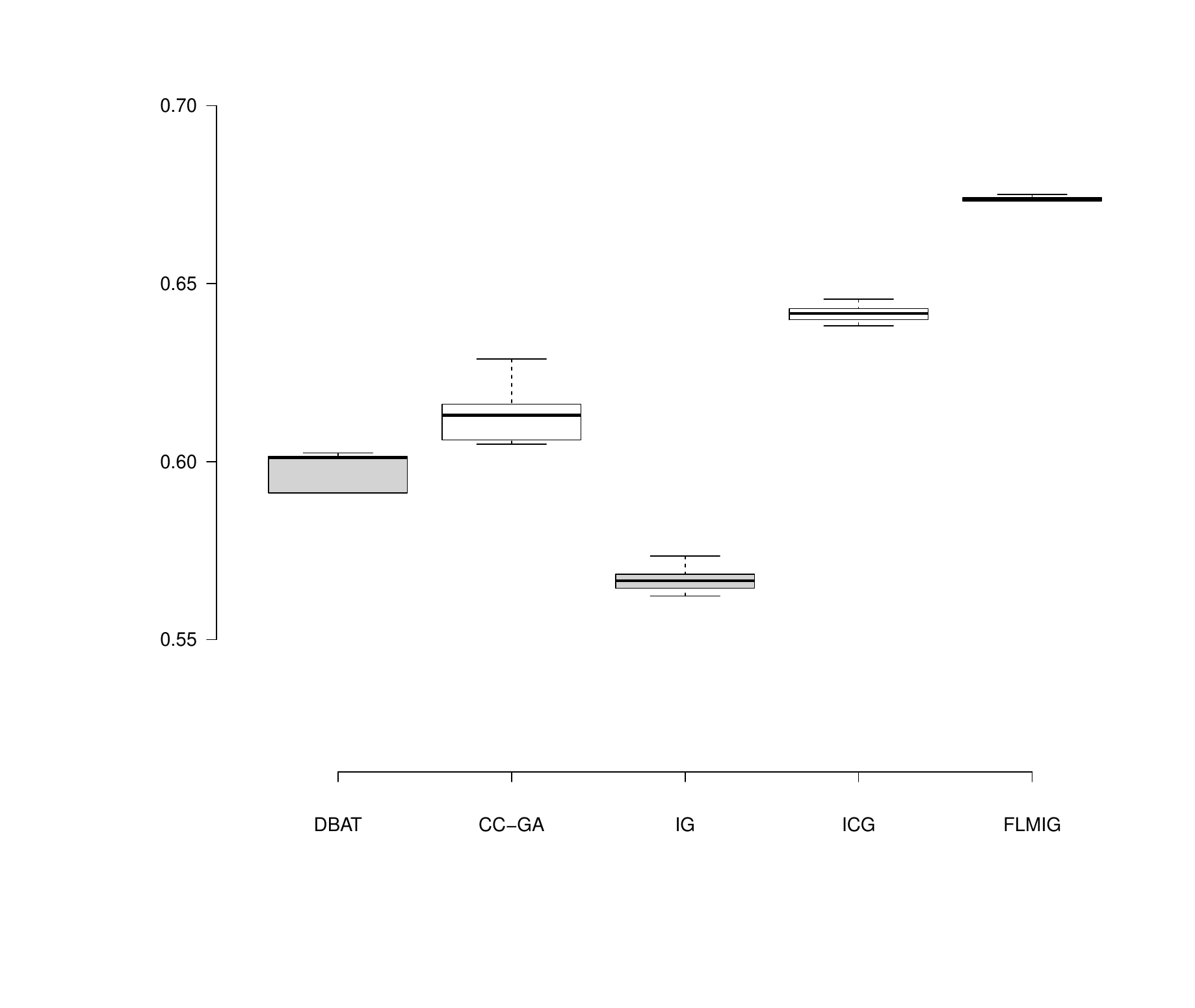} \quad
\caption{As-22jully06}
      \label{fig:sub1b}
    \end{subfigure}\quad
        \centering
    \begin{subfigure}[b]{.45\textwidth}
      \centering
    \includegraphics[scale=.25]{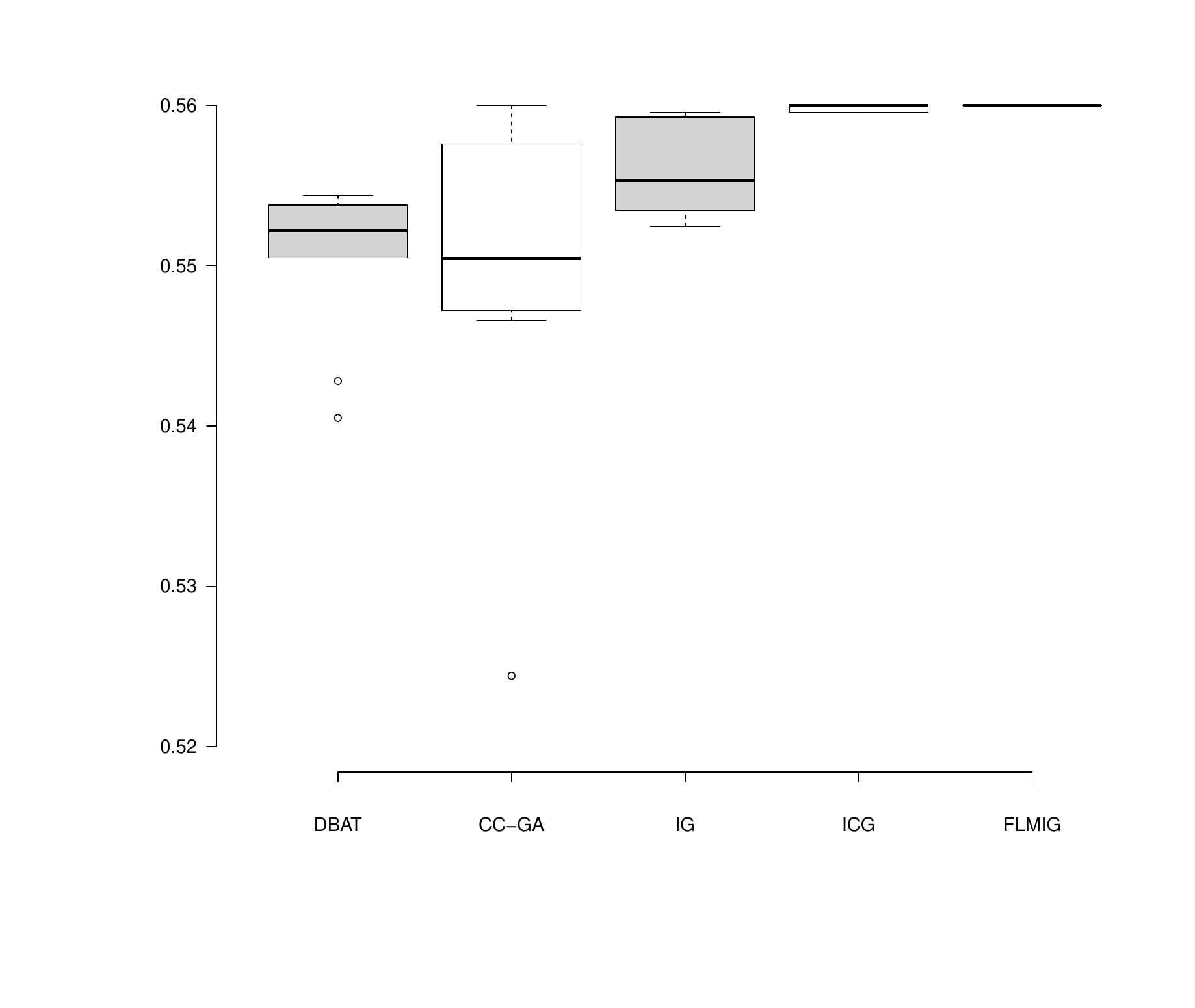} \quad
\caption{Lesmiss}
      \label{fig:sub1b}
    \end{subfigure}\quad
        \centering
    \begin{subfigure}[b]{.45\textwidth}
      \centering
    \includegraphics[scale=.25]{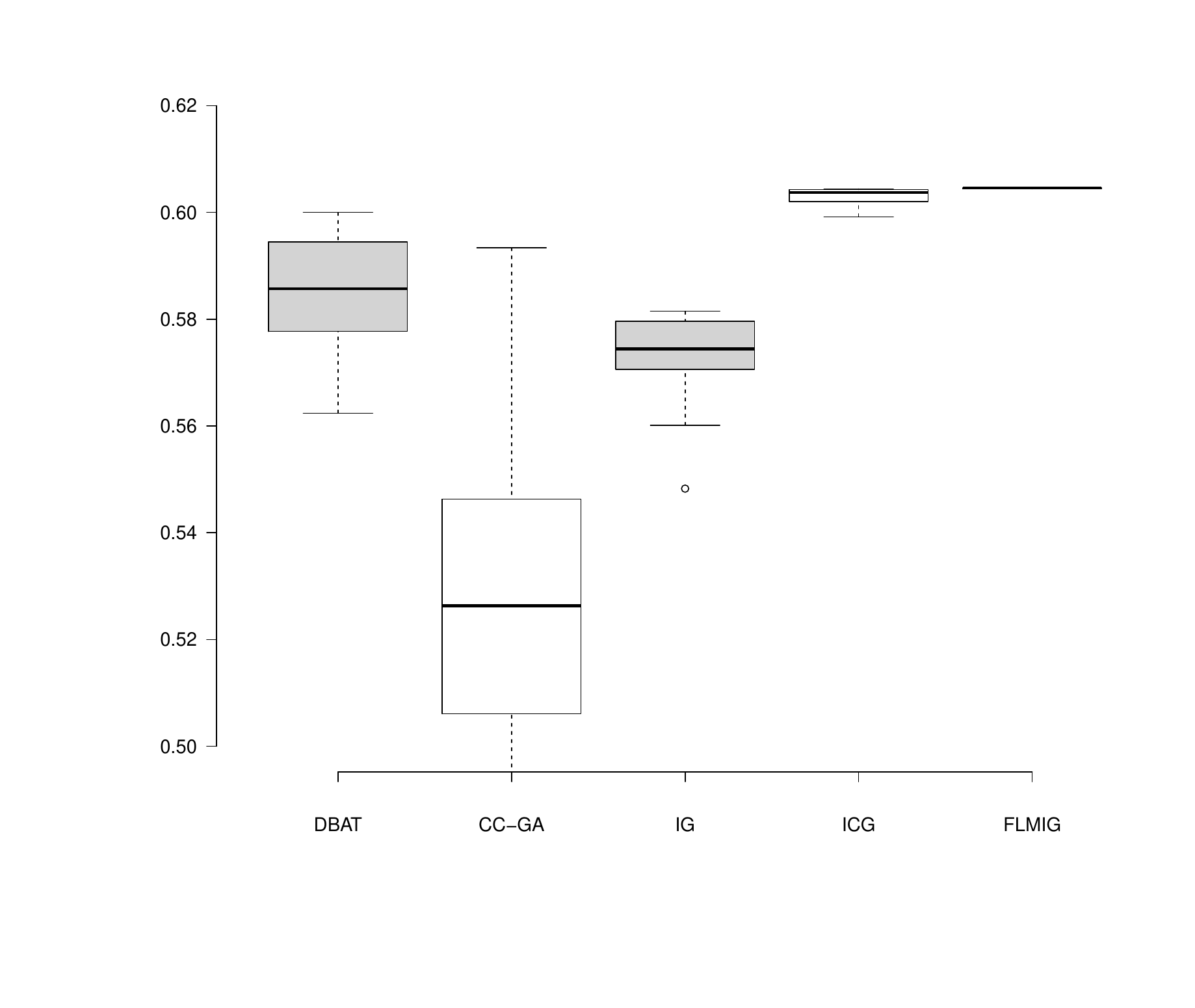} \quad
\caption{Football}
      \label{fig:sub1b}
    \end{subfigure}\quad
      \centering
    \begin{subfigure}[b]{.45\textwidth}
      \centering
    \includegraphics[scale=.25]{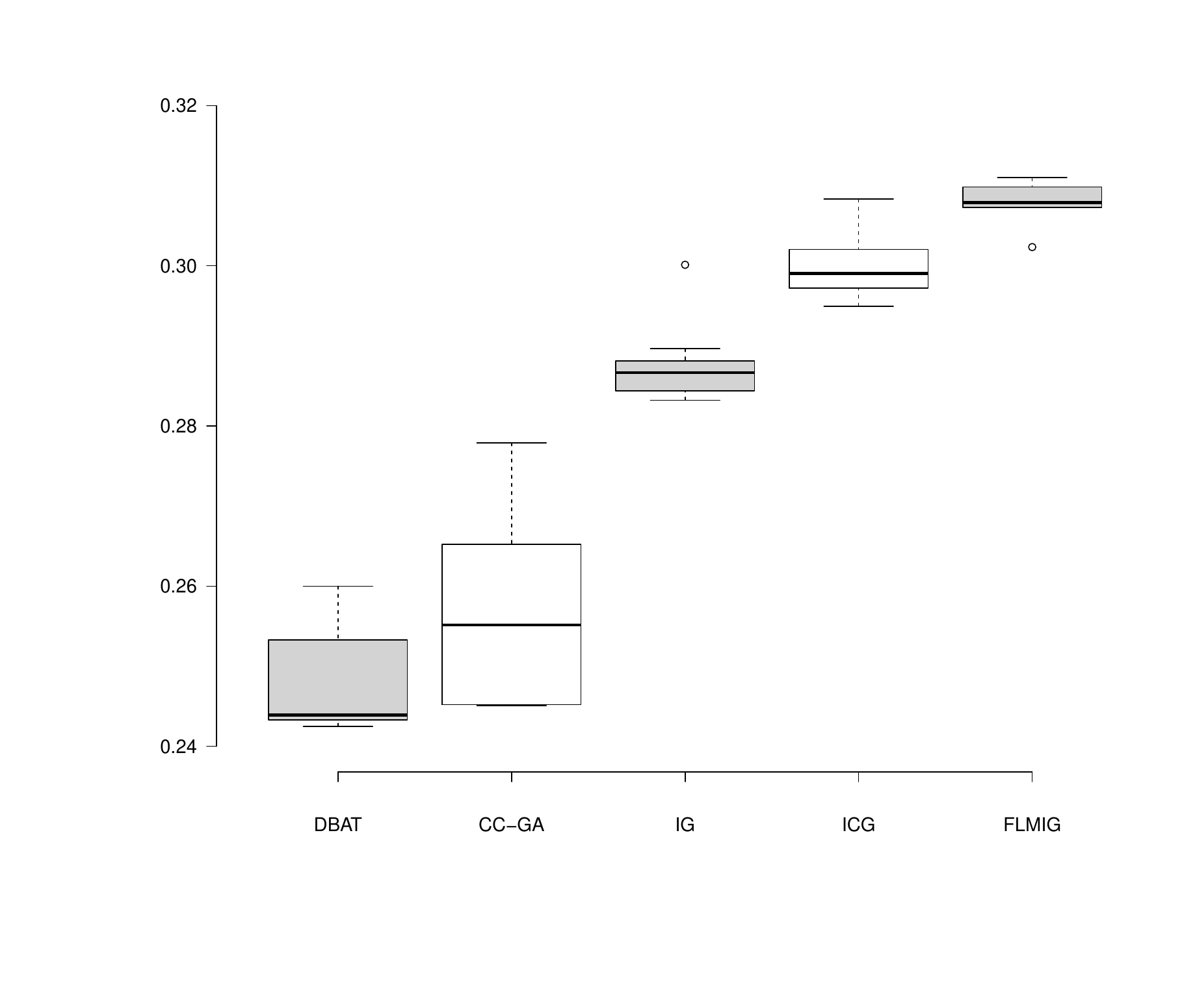} \quad
\caption{Adjnoun}
      \label{fig:sub1b}
    \end{subfigure}\quad
       \centering
    \begin{subfigure}[b]{.45\textwidth}
      \centering
    \includegraphics[scale=.25]{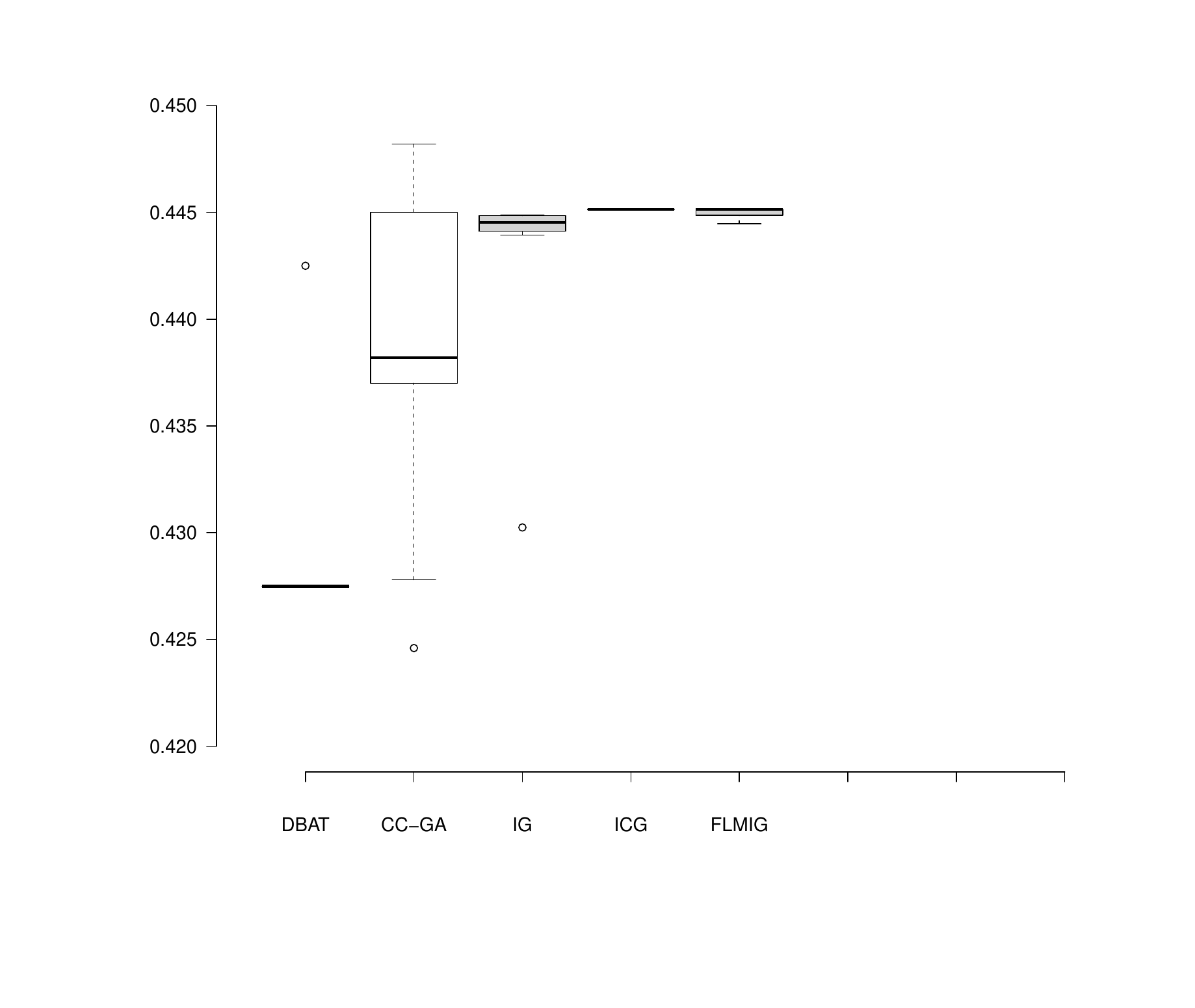} \quad
\caption{Jazz}
      \label{fig:sub1b}
    \end{subfigure}\quad
\caption{Experimental results of FLMIG and the compered algorithms on smaller and medium size real-world networks (2)}
\label{fig9}
  \end{minipage}\\[1em]
\end{figure}

\begin{figure}[H]
 \advance\leftskip-2cm
  \begin{minipage}{1.2\textwidth}
    \centering
    \begin{subfigure}[b]{.45\textwidth}
      \centering
    \includegraphics[scale=.25]{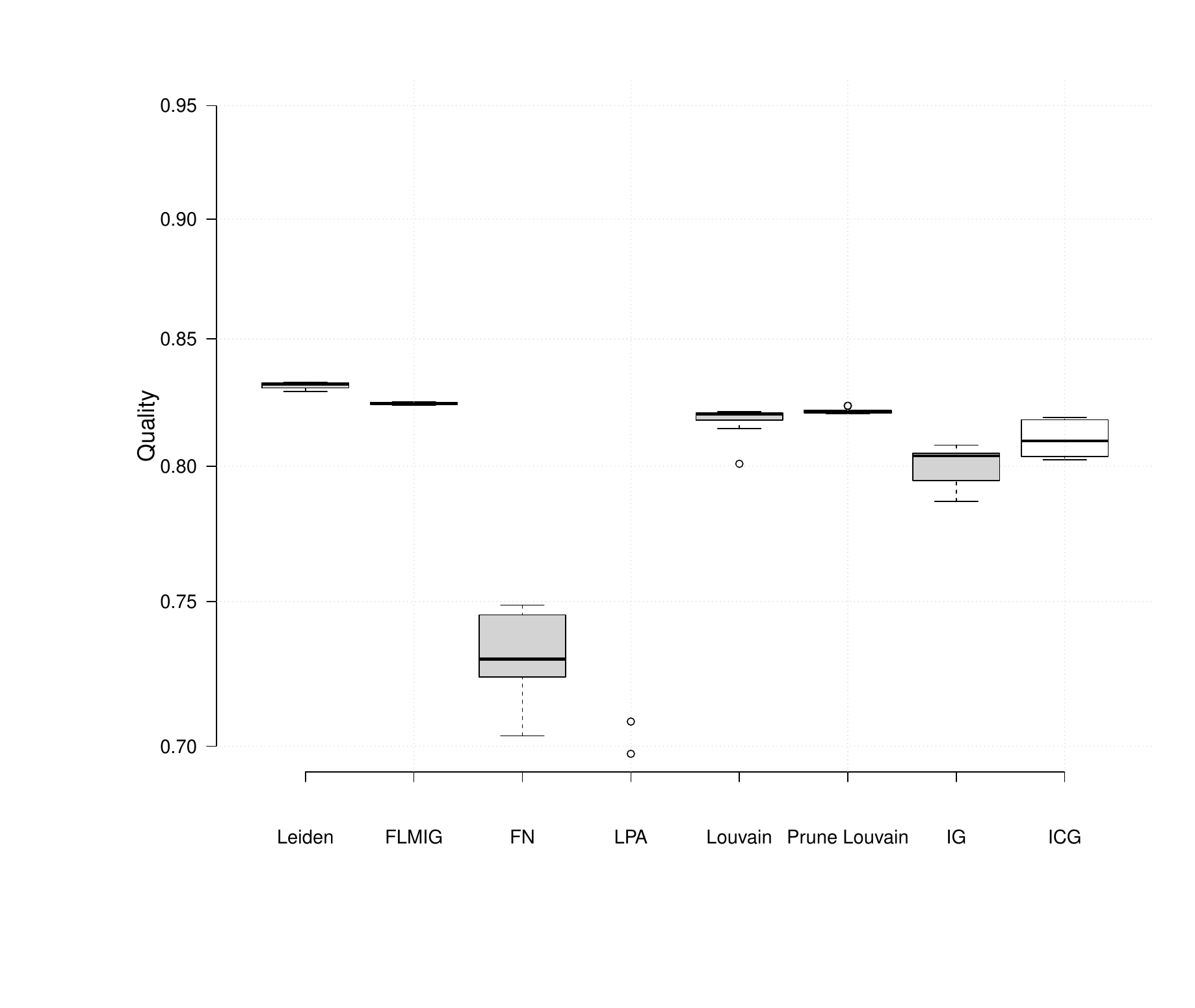} \quad
    \caption{Dblp}
      \label{fig:sub1a}
    \end{subfigure}\quad
    \centering
    \begin{subfigure}[b]{.45\textwidth}
      \centering
    \includegraphics[scale=.25]{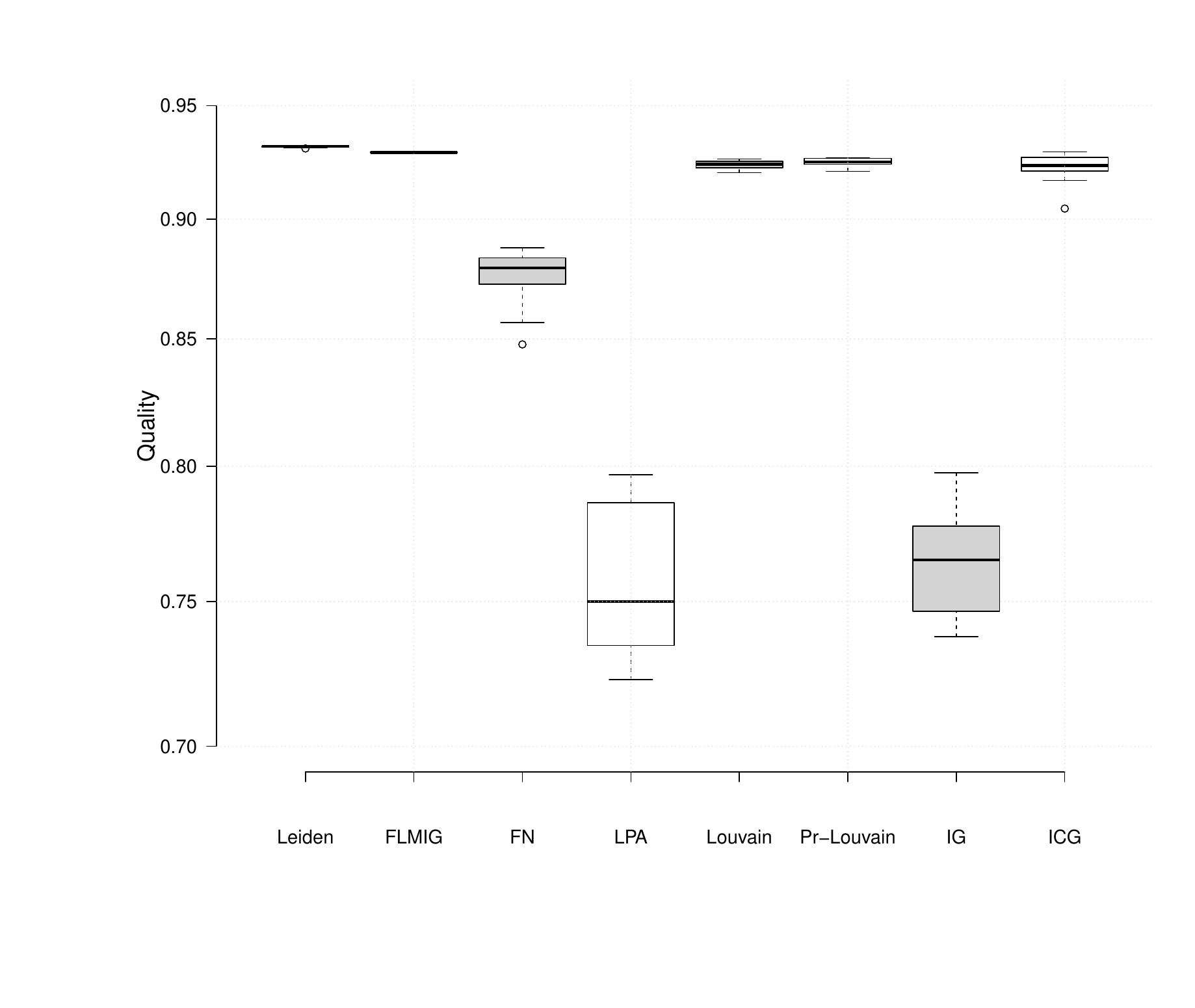} \quad
\caption{Amazon}
      \label{fig:sub1b}
    \end{subfigure}\quad
  
\caption{Experimental results of FLMIG and comparative world networks algorithms on large real-world networks}
\label{fig10}
  \end{minipage}\\[1em]
\end{figure}

\textcolor{blue}{ By comparing the suggested method FLMIG to other techniques on twelve real-world networks, we evaluated its performance. For each network, each algorithm was run ten times separately. The findings, displayed in figures Fig.  \ref{fig8},  Fig.  \ref{fig9}, Fig.   \ref{fig10} different datasets yield different results from each algorithm; some tendencies emerge. In general, datasets with high overall algorithmic success rates yield higher results for the ICG and FLMIG approaches, suggesting that under optimal conditions they could be more robust. The improved consistency with which CC-GA and DBAT operate across a range of datasets implies that they might be applicable in more contexts.}

\begin{figure}[t]
  \begin{minipage}{\textwidth}
    \centering
    \begin{subfigure}[b]{.4\textwidth}
      \centering
    \includegraphics[scale=.4]{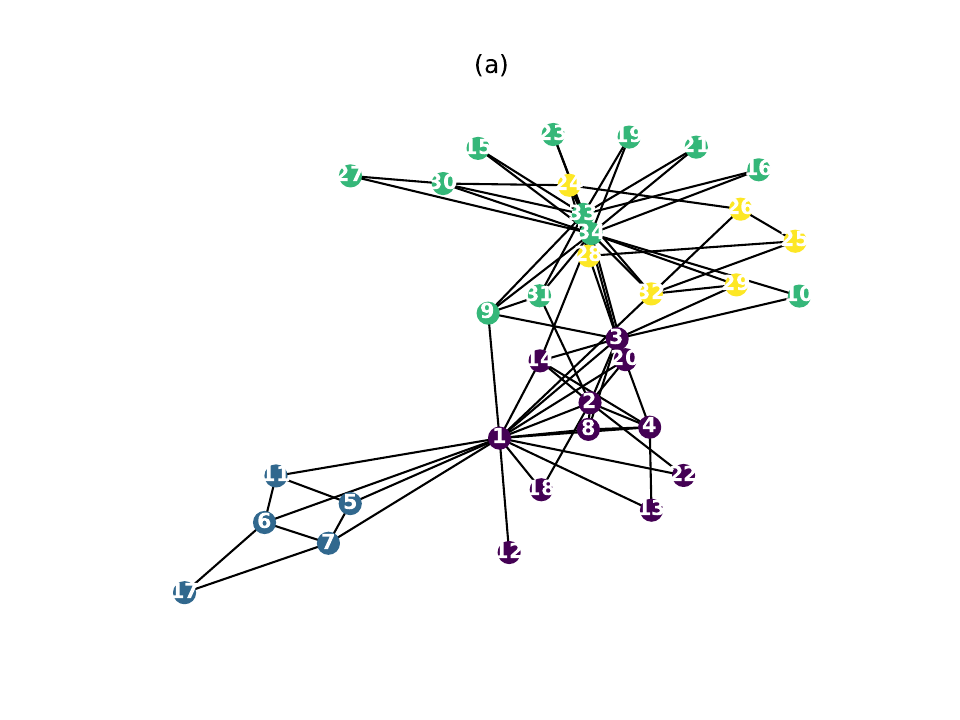} \quad
    \caption{Karate}
      \label{fig:sub1a}
    \end{subfigure}\quad
    \centering
    \begin{subfigure}[b]{.4\textwidth}
      \centering
    \includegraphics[scale=.4]{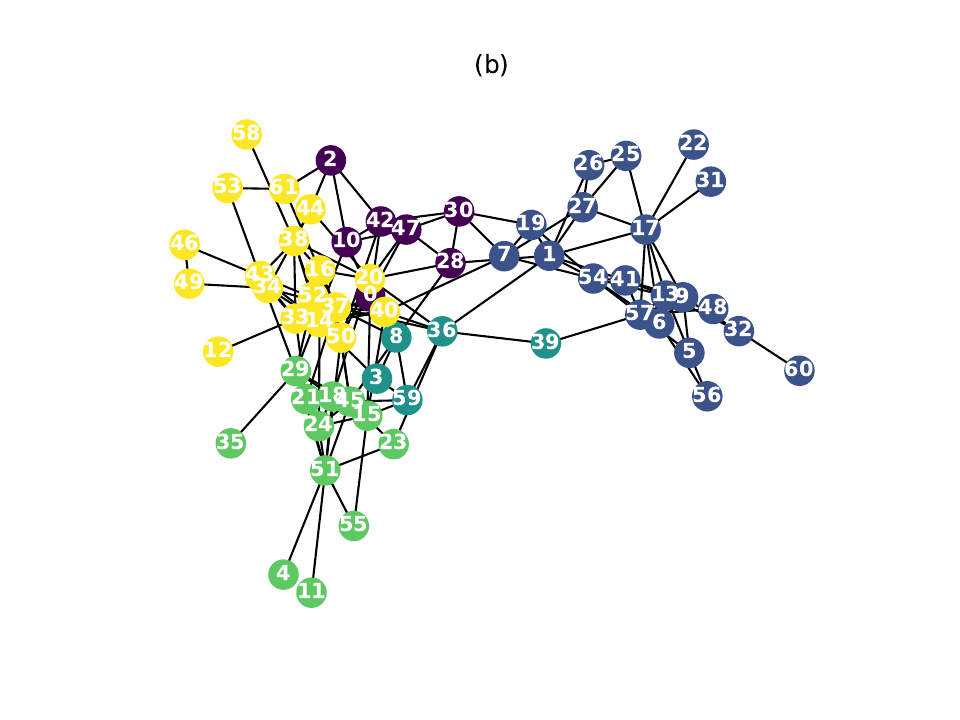} \quad
\caption{Dolphins}
      \label{fig:sub1b}
    \end{subfigure}\quad
        \centering
    \begin{subfigure}[b]{.4\textwidth}
      \centering
    \includegraphics[scale=.4]{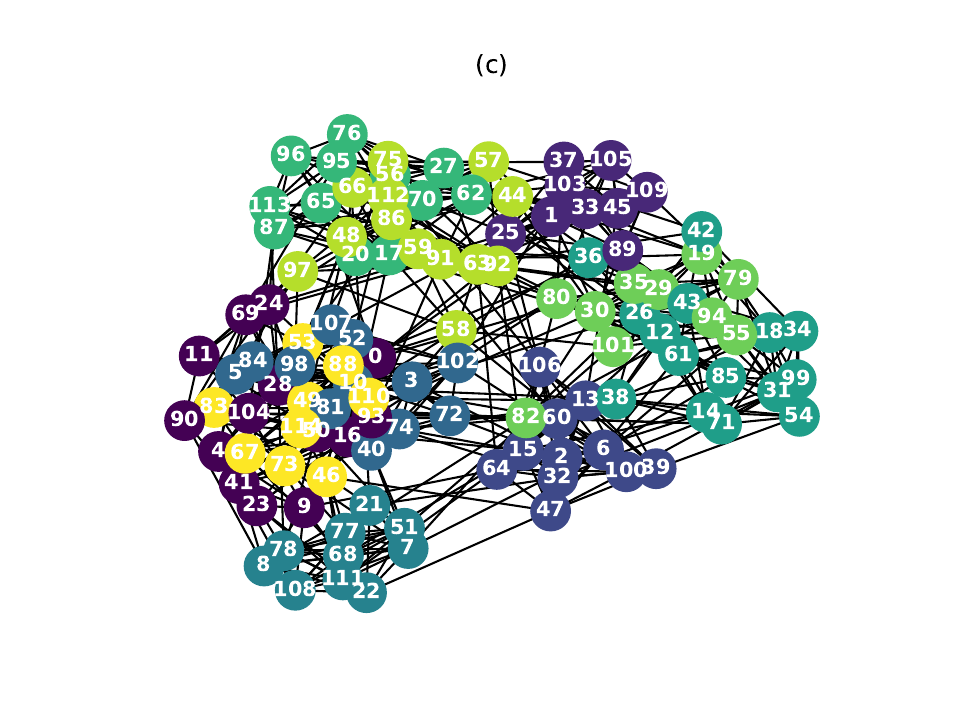}\quad
\caption{Football}
      \label{fig:sub1b}
    \end{subfigure}\quad
        \begin{subfigure}[b]{.4\textwidth}
      \centering
    \includegraphics[scale=.4]{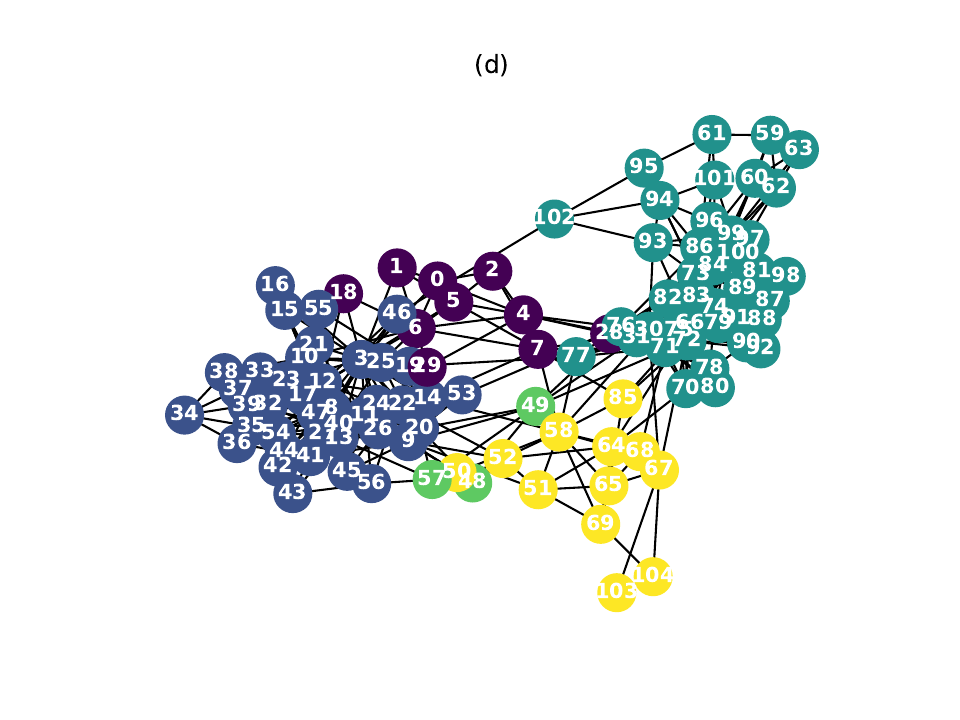}
\caption{Polbooks}
      \label{fig:sub1b}
    \end{subfigure}\quad
\caption{The detected communities on smaller size real-world networks}
\label{fig11}
  \end{minipage}\\[1em]
\end{figure}

Compared to other metaheuristics methods like ICG, DBAT-M, and CC-GA.  The FLMIG algorithm generally reach the best solutions with excellent stability and showed a shorter computation time in Table \ref{tab5},  especially in smaller and moderate-sized networks. Moreover, as Fig. \ref{fig11} illustrates, FLMIG demonstrated effectiveness in accurately identifying community structures in smaller networks with established ground realities.

Notably, the IG algorithm showed the fastest computational times; however, it was unable to attain appreciable modularity values, particularly in medium-sized networks like As-22July06 and PGP. Moreover, the integration of a local exploration with IG, as demonstrated in ICG, resulted in longer calculation times to achieve better results than IG alone. However, the proposed algorithm, FLMIG, performed better after integrating FLM with IG. For moderate-size networks, it outperformed ICG in terms of both solution quality and time efficiency.

We added four additional heuristics to the comparison, FN\cite{bib31}, Louvain\cite{bib11}, Louvain Prune \cite{bib24}, \textcolor{blue}{Leiden \cite{bib12}}, and LPA\cite{bib58}, in order to address the problem of metaheuristics methods like DBAT-M and CC-GA taking a long time to converge on large-scale networks. The empirical results, as shown in Fig. \ref{fig10}, show that FLMIG performs better than the current method in terms of stability on both big networks, DBLP and Amazon. The FLMIG method performs similarly to the Leiden algorithm, however, both algorithms surpass the IG, Louvain, prune Louvain, FN, and LPA algorithms.

\textcolor{blue}{ As previously stated, the community structure cannot be defined by the metaheuristic methods due to their high computing costs. In addition, FLIMG takes less computational time to define the community structure than the ICG, DBAT-M, and CC-GA metaheuristic methods. On the other hand, as table \ref{tab6} shows, the existing methods, Leiden, Louvain, Prune Louvain, consume a little computing time to locate the communities. But the best quality is obtained by Leiden and FLMIG}


\begin{table}[t]
\caption{\textcolor{blue}{Computational time in second(s) of FLMIG and the compared algorithms on smaller and medium size real-world networks}}\label{tab5} 
\begin{tabular}{ p{1.5cm} p{1.5cm} p{1.5cm} p{1.5cm} p{1.5cm} p{1.5cm} p{1.5cm}}
\toprule
Networks & Criterion & DBAT-M & CC-GA & IG & ICG & FLMIG \\
\midrule
\\[0.5pt]
& $Q_{best}$ & \textbf{0.4020} & \textbf{0.4198} & \textbf{0.4198} & \textbf{0.4198} & \textbf{0.4198} \cr
\multirow{1}{*}{\textbf{Karate}}& $Time(s)$ & 3.151 & 71.6 &  \textbf{0.07}   & 0.15 & 0.10
\\[5pt]
\hline
\\[0.5pt]
& $Q_{best}$ & 0.5277 & \textbf{0.5285} & \textbf{0.5285} & \textbf{0.5285}  &\textbf{0.5285}\cr
\multirow{1}{*}{\textbf{Dolphins}} & $Time(s)$ & 6.392 &   130.75   &   \textbf{0.1201}   &  0.5002   & 0.5129\\
[5pt]
\hline
\\[0.5pt]
& $Q_{best}$ & \textbf{0.6000} & \textbf{0.5943} & \textbf{0.6046} & \textbf{0.6046} & \textbf{0.6046}\cr
\multirow{1}{*}{\textbf{Football}}& $Time(s)$ & 16.02 & 357.12   & \textbf{0.3070} &  1.1823    & 0.9308\\
[5pt]

\hline
\\[0.5pt]
& $Q_{best}$ & 0.5235& 0.5272 & 0.5269 & 0.5272 & \textbf{0.5272}\cr
\multirow{1}{*}{\textbf{Polbooks}}& $Time(s)$ & 202.3 &  204.5  & 0.3031 & 1.017 & \textbf{0.7592}\\
[5pt]
\hline
\\[0.5pt]

& $Q_{best}$ & 0.5544 & \textbf{0.5600} & \textbf{0.5600} & \textbf{0.5600} & \textbf{0.5600}\cr
\multirow{1}{*}{\textbf{Lesmis}} & $Time(s)$ &   9.57 &  67.05  & \textbf{0.2133} & \ 0.6209 & 0.4832\\
[5pt]
\hline
\\[0.5pt]

& $Q_{best}$ & 0.2600 & 0.2779 & 0.3007 & 0.3100 & \textbf{0.3130}\cr 
\multirow{1}{*}{\textbf{Adjnoun}}& $Time(s)$ & 10.94 & \  48.52   & 0.2479& \ 1.3788 & \textbf{0.9054}\\
[5pt]

\hline
\\[0.5pt]
& $Q_{best}$ & 0.4425 & 0.4420 & 0.4446 & \textbf{0.4451} & \textbf{0.4451} \cr 
\multirow{1}{*}{\textbf{Jazz}}& $Time(s)$ & 35.55 & 176.14 & 0.9528 &  4.3805 & \textbf{3.377}\\
[5pt]

\hline
\\[0.5pt]
& $Q_{best}$ & 0.4036 & 0.4024    & 0.4320 & 0.4452 & \textbf{0.4558} \cr 
\multirow{1}{*}{\textbf{Metabolic} }& $Time(s)$ & 81.80 &   168.5 & 1.075 & 6.4087 & \textbf{4.01} \\
[5pt]
\hline
\\[0.5pt]
& $Q_{best}$ & 0.9274 & 0.9584 & 0.9286 & 0.9363 & \textbf{0.9599} \cr 
\multirow{1}{*}{\textbf{Netscience} }& $Time(s)$ & 668.24 & 1465.1 &  2.18 & 8.93 & \textbf{5.91}\\
[5pt]

\hline
\\[0.5pt]
& $Q_{best}$ & 0.8070 & 0.8523 & 0.7689 & 0.8000 & \textbf{0.8850} \cr 
\multirow{1}{*}{\textbf{PGP} }& $Time(s)$ &  1975.8 & 1915 & 38.36 & 180.20 & \textbf{77.75}\\
[5pt]

\hline
\\[0.5pt]
& $Q_{best}$ & 0.6014 & 0.6142 & 0.6250 & 0.6454 & \textbf{0.6757} \cr 
\multirow{1}{*}{\textbf{As-22jully06} }& $Time(s)$ & 25063 & 24250 & 117.31 &  740.03 & \textbf{300.05}\\
\\[5pt]

\bottomrule

\end{tabular}
\end{table}

\begin{figure}[t]
  \begin{minipage}{1.1\textwidth}
    \centering
    \begin{subfigure}[b]{.45\textwidth}
      \centering
    \includegraphics[scale=.4]{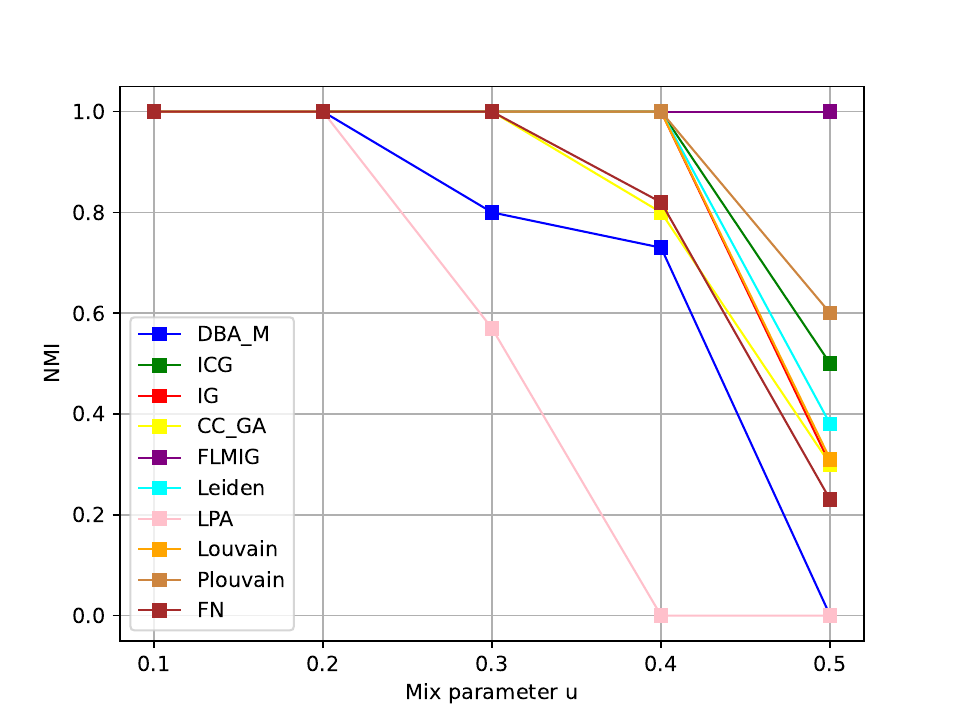} \quad
    \caption{GN benchmark}
      \label{fig:sub1a}
    \end{subfigure}\quad
    \centering
    \begin{subfigure}[b]{.45\textwidth}
      \centering
    \includegraphics[scale=.4]{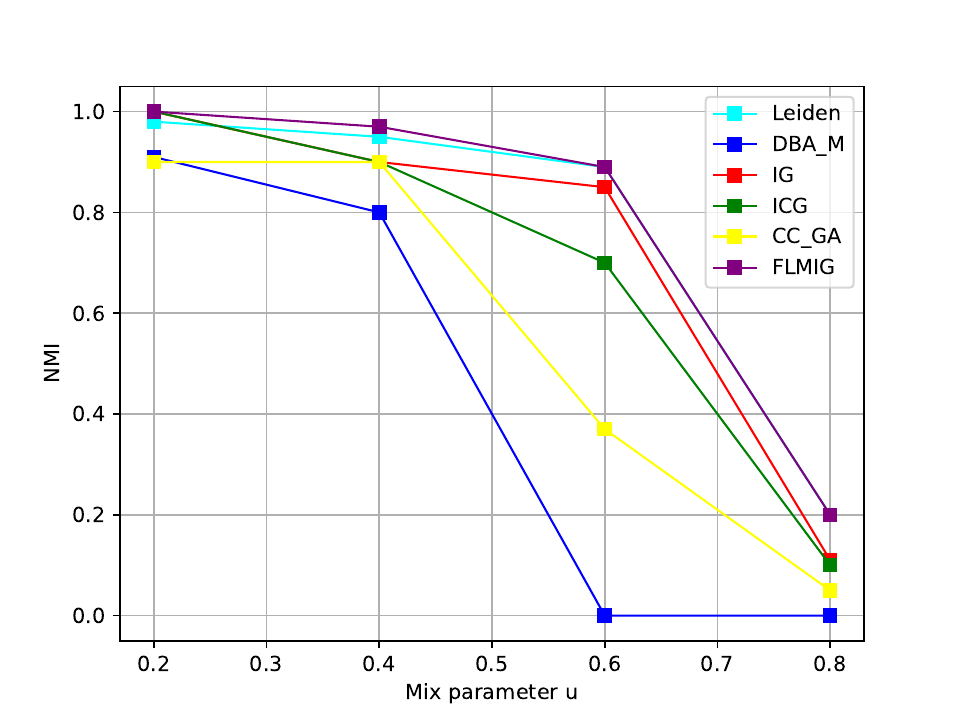} \quad
\caption{LFR(a) benchmark}
      \label{fig:sub1b}
    \end{subfigure}\quad
        \centering
    \begin{subfigure}[b]{.45\textwidth}
      \centering
    \includegraphics[scale=.4]{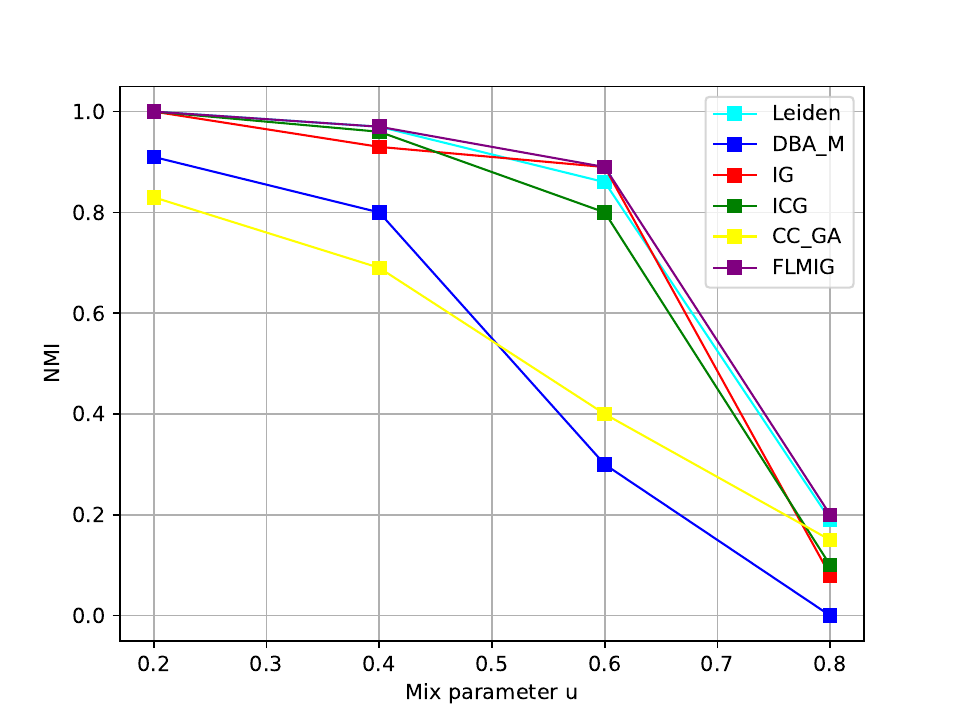}\quad
\caption{LFR(b) benchmark}
      \label{fig:sub1b}
    \end{subfigure}\quad
\caption{The maximum NMI achieved by DBAT-M, CC-GA, IG ,ICG and FLMIG algorithms}
\label{fig12}
  \end{minipage}\\[1em]
\end{figure}

\begin{figure}[t]
  \begin{minipage}{1.1\textwidth}
    \centering
    \begin{subfigure}[b]{.45\textwidth}
      \centering
    \includegraphics[scale=.4]{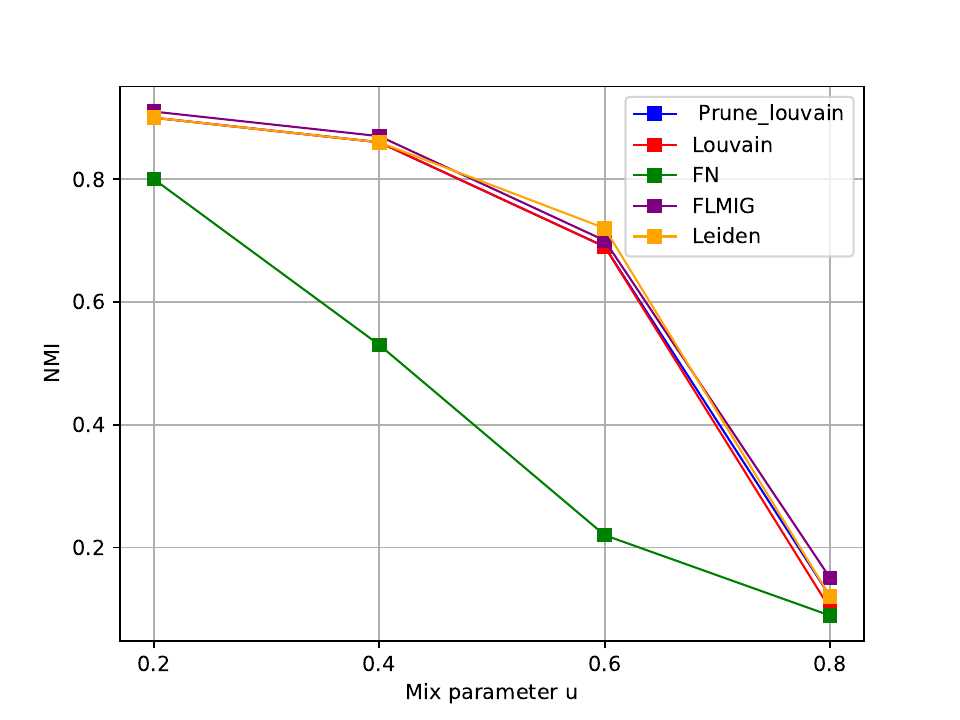} \quad
    \caption{LFR benchmark n =10000}
      \label{fig:sub1a}
    \end{subfigure}\quad
    \centering
    \begin{subfigure}[b]{.45\textwidth}
      \centering
    \includegraphics[scale=.4]{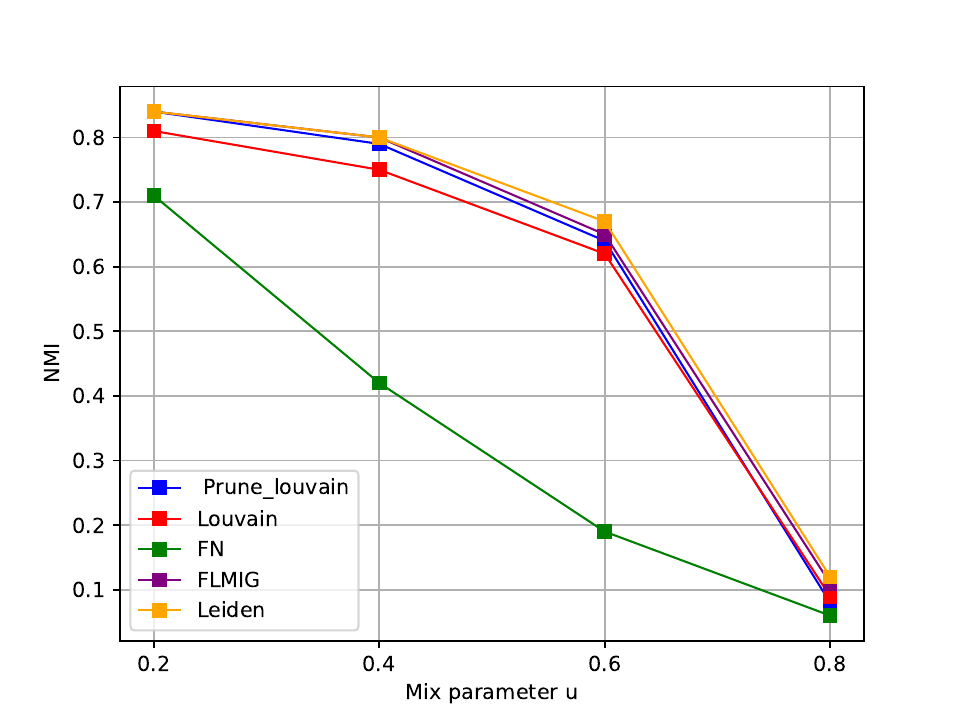} \quad
\caption{LFR benchmark n =100000}
      \label{fig:sub1b}
    \end{subfigure}\quad

\caption{The maximum NMI achieved by FN, LPA, Leiden, Louvain, Louvain Prune and FLMIG algorithms}
\label{fig13}
  \end{minipage}\\[1em]
\end{figure}

\begin{table}[t]
 \advance\leftskip-2cm

\caption{ \textcolor{blue}{Computational time in second(s) of FLMIG and the compared algorithms on large size real-world networks}}\label{tab6}%
\begin{tabular}{ p{1cm} p{1cm} p{1cm} p{1cm} p{1cm} p{1cm} p{1cm} p{1cm} p{1cm}}
\toprule
Networks  & Criterion & Leiden & FN  & IG & ICG & Louvain & Prune Louvain & FLMIG \\
\midrule
\\[0.5pt]

 & $Q_{best}$ & \textbf{0.8329 }& 0.7445 &  0.8047 & 0.8179 &  0.8190  & 0.8232 & 0.8250 \cr 
\multirow{1}{*}{\textbf{Dblp} }& $Time(s)$ & \textbf{40.15 }&  3910.451    & 2000.18 & 4900.6209  &100.7108 &90.88 &2500.6209\\
[5pt]

\hline
\\[0.5pt]

& $Q_{best}$ & \textbf{0.933} &   0.8795     &  0.7775   & 0.929 & 0.9233 & 0.9259 & 0.9296 \cr 
\multirow{1}{*}{\textbf{Amazon} }& $Time(s)$ &  \textbf{15.34} &   3500.78   &   1900.23   & 4600.62 & 46.71 &40.9 & 2400.6509\\
[5pt]

\bottomrule

\end{tabular}
\end{table}

\subsection{ Performance on synthetic networks}\label{subsec3}
We used synthetic networks with predetermined community structures for our evaluation, and the key evaluation criterion was the normalized mutual information (NMI) metric.

In relation to the GN benchmark performance, we carried out 20 separate iterations of all the algorithms being tested for each GN network. The findings, illustrated in Fig \ref{fig12}, indicate that FLMIG accurately recognized the appropriate divisions in all GN networks, as evidenced by the NMI values attaining a value of one.

\textcolor{blue}{Nevertheless, the IG ,ICG , Louvain, Louvain Prune and Leiden algorithms faced a constraint in precisely identifying the optimal partitions, especially when the mixing parameter $u$ above 0.4. Similarly, the FN,CC-GA algorithms encountered difficulty in detecting the community structure after the value of $u$ exceeded 0.3. The DBAT-M, LPA methods encountered difficulties in accurately identifying partitions for values of $u$ beyond 0.3, and its performance significantly declined when $u$ surpassed 0.4}

Based on this research, it is evident that our FLMIG method demonstrates a substantial advantage over these metaheuristics in the GN benchmark tests. This highlights its superior capacity to discover community structures, even in difficult circumstances.   

\subsection{ Performance on the LFR benchmark}\label{subsec4}
We performed 20 separate trials for each LFR network to evaluate all the metaheuristics being compared. The results, displayed in Fig \ref{fig12}, demonstrate that FLMIG excels at identifying community structures that closely align with the real divisions, surpassing other algorithms in performance.

In the scenario of LFR benchmark type (A), as shown in Fig \ref{fig12}, all metaheuristics exhibited the capacity to precisely identify \textcolor{blue}{a good community structure quality for networks with a mixing parameter $u$ below 0.4}. FLMIG distinguishes itself by accurately identifying the appropriate partitions for values of $u$ that are less than 0.6.

Concerning the LFR benchmark type (B), as demonstrated in Fig \ref{fig12}, FLMIG outperforms all other metaheuristics that were compared. It should be emphasized that as the value of $u$ increases, the task of accurately recognizing the community structure gets more difficult. However, FLMIG routinely demonstrates superior performance in reliably identifying valid partitions, even in challenging circumstances.
\textcolor{blue}{Regarding the large LFR benchmark, as shown in Fig \ref{fig13}, in the networks with n = $10^4$, FLMIG outperforms all the compared algorithms except where $u$ = 0.6, the Leiden outperforms the existing approaches. In the networks with n = $10^5$, the performance of FLMIG is closer to the Leiden algorithm with different mix parameters $u$; as a result, the FLIMG proves combativeness in synthetics networks with Leiden concerning the $NMI$ metric}.

\section{Conclusion}\label{sec5}

\textcolor{blue}{The Fast Local Move Iterated Greedy (FLMIG) algorithm was developed in this research endeavor as an innovative strategy to improve the detection of communities in intricate networks. By combining fast local move heuristics with an iterated greedy framework, FLMIG substantially enhances upon existing approaches by enabling more rapid convergence and precise community partitioning.}

\textcolor{blue}{The experimental findings, obtained from both real-world and synthetic networks, indicate that FLMIG surpasses conventional approaches such as the Louvain and Leiden algorithms in terms of computational efficiency and modularity. Moreover, FLMIG demonstrates resilient performance across a wide range of network scales and types. This demonstrates the adaptability and efficacy of the algorithm in managing networks that are dynamic and expansive in size.}

\textcolor{blue}{FLMIG's efficacy is substantially enhanced by its innovative components, including the enhanced prune Louvain algorithm and the incorporation of random neighbor movements. These characteristics guarantee that the algorithm effectively manages the exploration and exploitation stages, resulting in the detection of communities of superior quality.}

\textcolor{blue}{Subsequent investigations may delve into the utilization of FLMIG in networks comprising overlapping community structures, as well as the potential of parallel processing to further diminish computation duration. Moreover, by extending FLMIG to encompass temporal and multiplex networks, among other intricate systems, its utility and influence could be significantly expanded in the domain of network science.}

\textcolor{blue}{This study not only contributes to the existing body of knowledge on community detection algorithms but also offers a practical and efficient instrument for practitioners and researchers in the field of network analysis across multiple academic disciplines.}
\backmatter

\bmhead{Acknowledgments}

This project is supported by the Algerian Directorate General of Scientific Research and Technological Development (DGSRTD).

\bibliographystyle{plain}
\bibliography{sn-bibliography.bib}

\end{document}